\documentclass[12pt,preprint]{aastex}
\shorttitle{Corotation torque in an adiabatic disk}
\shortauthors{Baruteau and Masset}

\begin{document}
\title{On the corotation torque in a radiatively inefficient disk}

\author{C. Baruteau\altaffilmark{1} and F. Masset\altaffilmark{2}}
\affil{Laboratoire AIM, CEA/DSM - CNRS - Universit{\'e} Paris Diderot,
  DAPNIA/Service d'Astrophysique, CEA/Saclay, 91191 Gif/Yvette Cedex,
  France; clement.baruteau@cea.fr; fmasset@cea.fr}
\altaffiltext{1}{Send offprint request to clement.baruteau@cea.fr.}
\altaffiltext{2}{Also at IA-UNAM, Ciudad Universitaria, Apartado
  Postal 70-264, Mexico D.F. 04510, Mexico.}

\begin{abstract}
  We consider the angular momentum exchange at the corotation
  resonance between a two-dimensional gaseous disk and a uniformly
  rotating external potential, assuming that the disk flow is
  adiabatic.  We first consider the linear case for an isolated
  resonance, for which we give an expression of the corotation torque
  that involves the pressure perturbation, and which reduces to the
  usual dependence on the vortensity gradient in the limit of a cold
  disk. Although this expression requires the solution of the
  hydrodynamic equations, it provides some insight into the dynamics
  of the corotation region. In the general case, we find an additional
  dependence on the entropy gradient at corotation.  This dependence
  is associated to the advection of entropy perturbations. These are
  not associated to pressure perturbations. They remain confined
  to the corotation region, where they yield a singular contribution
  to the corotation torque.  In a second part, we check our torque
  expression by means of customized two-dimensional hydrodynamical
  simulations. In a third part, we contemplate the case of a planet
  embedded in a Keplerian disk, assumed to be adiabatic. We find an
  excess of corotation torque that scales with the entropy gradient,
  and we check that the contribution of the entropy perturbation to
  the torque is in agreement with the expression obtained from the
  linear analysis. We finally discuss some implications of the
  corotation torque expression for the migration of low mass planets
  in the regions of protoplanetary disks where the flow is radiatively
  inefficient on the timescale of the horseshoe U-turns.
\end{abstract}

\keywords{accretion, accretion disks --- hydrodynamics --- methods:
  numerical --- planetary systems: formation --- planetary systems:
  protoplanetary disks}

\section {Introduction}
\label{sec:intro}

It is known since the early eighties that low mass planetary objects
(that is, up to a few Earth masses) embedded in protoplanetary gaseous
disks should undergo a fast decay towards their central object, on
timescales much shorter than the lifetime of the disk. This process,
known as type~I migration, has constituted for a long time a critical
stage for the theory of giant planet formation. While it may account
for the discovery of close-in extrasolar planets, with orbital periods
of a few days, it renders problematic the build up of giant planet
cores at distances of their central stars of several astronomical
units. Most published studies of the tidal interaction of low mass
objects with their parent disk have used either a barotropic
assumption (such as a polytropic equation of state), or a locally
isothermal equation of state. All these studies, whether analytical or
numerical, confirmed the vigorous tidal interaction of the planet with
the disk, leading to its inward migration on short timescales.

There has been some exceptions to such assumptions: \citet{mota03}
considered the case of a planet interacting with an optically thin
disk, in the shearing sheet approximation, and found that radiative
effects can significantly alter the one-sided torque between the
planet and the disk. More recently, \citet{pm06} (hereafter PM06) have
performed global, high resolution 3D calculations with nested grids
that include radiative transfer. For the setup that they considered,
they found that the total torque exerted by the disk on the planet
increases with the disk opacity.  For sufficiently large values of the
opacity (and in the limit case of an adiabatic flow, corresponding to
an infinite opacity), they find that the total torque on the planet is
positive.  This result is of great importance, as it potentially
solves the lingering problem of type~I migration. PM06 identified the
existence of a hot, underdense part of the co-orbital region lagging
the planet, which accounted for the torque excess that they
measured. The present work corresponds to an attempt to further
investigate this topic, so as to identify the physical mechanism
responsible for these effects. For this purpose, we consider a more
restricted situation, namely two-dimensional adiabatic flows.

This paper is organized as follows. In section~\ref{sec:not} we set up
the problem and define the notation.  We then present an analysis of
the corotation torque in an adiabatic disk in the linear regime, at an
isolated resonance, at section~\ref{sec:isores}.  Our original
motivation for the study of the linear regime was that PM06 found that
the total torque reverses in a radiatively inefficient disk both for a
$5\;M_\oplus$ and a $0.5\;M_\oplus$ planet, which pointed out that the
effect is likely a linear one. In section~\ref{sec:numiso}, we check
by means of customized two-dimensional hydrodynamical simulations
involving an isolated resonance the torque expression found in
section~\ref{sec:isores}. In section~\ref{sec:planet}, we turn to the
case of a planet embedded in an adiabatic disk, for which we check
that there is an excess of corotation torque that scales with the
entropy gradient.  We also check in this section that the torque
excess corresponds to the sum of the linear contributions of all
co-orbital corotation resonances, for a sufficiently small planet
mass. We discuss the implications of the modified corotation torque
expression for the issue of planet--disk tidal interactions, and we
suggest further research on this topic in section~\ref{sec:discu}. We
sum up our results in section~\ref{sec:conc}.

\section{Setup and notation}
\label{sec:not}

We consider an inviscid, radiatively inefficient (that is to say, for
our purposes, adiabatic) two-dimensional disk.  In order to avoid
corotation torque issues, we shall consider either a potential slowly
turned on (sections~\ref{sec:isores} and~\ref{sec:numiso}), or early
stages after the introduction of a planet
(section~\ref{sec:planet}). The unperturbed state of the disk
corresponds to a rotational equilibrium between the gravitational
force of the central object, the pressure gradient and the centrifugal
force.  The unperturbed state is axisymmetric. The disk rotates with
the angular speed $\Omega(r)$, where $r$ is the distance to the
central object. We denote by $p$ the pressure, $\Sigma$ denotes the
surface density, $u$ and $v$ respectively the radial and azimuthal
velocities, $\varphi$ the azimuthal angle. We denote by a ``$0$''
subscript the unperturbed quantities, and with a ``$1$'' subscript the
perturbed ones.  For instance, $p(r,\varphi) = p_0(r)+p_1(r,\varphi)$.
We shall essentially consider disks in which the unperturbed pressure
and density are power laws of the radius, respectively with index
$\lambda$ and $\sigma$:
\begin{eqnarray}
  p_0(r)& \propto& r^{-\lambda}\\
  \Sigma_0(r)&\propto& r^{-\sigma}.
\end{eqnarray}
We shall make use of the two Oort's constants:
\begin{equation}
  A=\frac 12r\frac{d\Omega}{dr},
\end{equation}
which scales with the local shear in the flow, and:
\begin{equation}
  B=\frac{1}{2r}\frac{d(r^2\Omega)}{dr}=\Omega+A,
\end{equation}
which is half the vertical component of the flow vorticity, and which
is also $(2r)^{-1}$ times the radial derivative of the specific
angular momentum.  We will also use the epicyclic frequency
$\kappa=(4\Omega B)^{1/2}$.

\section{Linear analysis at an isolated resonance}
\label{sec:isores}

\subsection {Basic equations}
\label{sec:basic}

We study the linear response of the disk to a perturbing
non-axisymmetric potential $\Phi
(r,\varphi)=\Phi_m(r)\cos[m(\varphi-\Omega_pt)]$.  The perturbing
potential rotates at constant angular velocity $\Omega_p$.  In the
inertial frame, the linearized Euler equations of the disk are:
\begin{equation}
  \frac{\partial u_1}{\partial t} + 
  \Omega\frac{\partial u_1}{\partial \varphi} - 2\Omega v_1
  = -\frac{\partial \Phi}{\partial r} - 
  \frac{1}{\Sigma_0}\frac{\partial p_1}{\partial r} + 
  \frac{\Sigma_1}{\Sigma_0^2}\frac{\partial p_0}{\partial r}
\label{eqnnsr}
\end{equation}
and
\begin {equation}	
  \frac{\partial v_1}{\partial t} + \Omega\frac{\partial v_1}{\partial
    \varphi} + \frac{\kappa^2}{2\Omega} u_1 =
  -\frac{1}{r}\frac{\partial}{\partial\varphi} \left(\Phi +
    \frac{p_1}{\Sigma_0} \right).
\label{eqnnst}
\end {equation}
The linearized continuity equation is:
\begin {equation}
  \frac{\partial \Sigma_1}{\partial t} + \Omega\frac{\partial
    \Sigma_1}{\partial \varphi} + \frac{1}{r}\frac{\partial}{\partial
    r} \left( r\Sigma_0 u_1\right) +
  \frac{1}{r}\frac{\partial}{\partial \varphi}\left( \Sigma_0
    v_1\right) = 0.
\label{eqncont}
\end {equation}
We refer to the quantity $S=p \Sigma^{-\gamma}$ as the gas entropy,
where $\gamma$ is the adiabatic index.  The energy equation is
equivalent in our case to the conservation of the gas entropy.  The
linearized conservation of the entropy along a fluid element path
reads
\begin {equation}
  \frac{\partial S_1}{\partial t} + \Omega\frac{\partial S_1}{\partial
    \varphi} + u_1 \frac{\partial S_0}{\partial r} = 0,
\label{eqnent}
\end {equation}
where $S_1 = S_0 (p_1 / p_0 - \gamma\Sigma_1 / \Sigma_0)$.  We
furthermore assume that the gas is described by an ideal equation of
state so that $p_0$ and $\Sigma_0$ are connected by $p_0 = \Sigma_0
c_s^2 / \gamma$, $c_s$ being the adiabatic sound speed.

We assume a perturbation of the form $x_{1,m}(r) \exp
(im\{\varphi-\Omega_p t\})$ where $x_1$ stands for any perturbed
quantity of the flow\footnote{We drop the subscript $m$ in
  $x_{1,m}(r)$ to improve legibility.}.  We note $\Delta\omega =
m(\Omega_p - \Omega)$ and we use the prime notation to denote
$\partial / \partial r$.  Eq.~(\ref{eqnent}) can be recast as:
\begin{equation}
  \Sigma_1 = \frac{p_1}{c_s^2} + \frac{i{\cal S}\Sigma_0 u_1}{r \Delta\omega}.
\label{s1}
\end{equation}
Combining Eqs.~(\ref{eqnnsr}), (\ref{eqnnst}) and~(\ref{s1}) we are
led to:
\begin {equation}
  \Sigma_0 u_1 = i{\cal F}\left[ \frac{\Delta\omega}{\Omega} \left\{
      \left( \Phi+\Psi \right)^{'} -\frac{{\cal S}}{r}\Psi \right\} -
    \frac{2m}{r}(\Phi + \Psi) \right]
\label{s0u1}
\end {equation}
and
\begin {equation}
  \left. \Sigma_0 v_1 \right. = {\cal F}\left[
    \frac{\kappa^2}{2\Omega^2} \left\{ \left( \Phi+\Psi \right)^{'} -
      \frac{{\cal S}}{r}\Psi \right\} \right. \nonumber \left. -
    \frac{m}{r} \left\{ \frac{\Delta\omega}{\Omega} + {\cal S}{\cal
        P}\frac{c_s^2/r^2}{\Delta\omega \Omega} \right\}\left(
      \Phi+\Psi \right) \right],
\label{s0v1}
\end {equation}
where ${\cal S}$ and ${\cal P}$ are given by
\begin{equation} 
{\cal S} = \frac{1}{\gamma}\frac{d\ln S_0}{d \ln r}
\label{cals}
\end{equation}
and
\begin{equation} 
{\cal P} = \frac{1}{\gamma}\frac{d\ln p_0}{d \ln r},
\label{calp}
\end{equation}
where $\Psi$ is defined as
\begin{equation}
  \Psi=p_1/\Sigma_0
\end{equation}
and where ${\cal F}$ is defined by
\begin{equation}
  {\cal F} = \frac{\Sigma_0 \Omega}{D},
  \label{calf}
\end{equation}
with $D=\kappa^2-\Delta\omega^2 - {\cal S}{\cal P}c_s^2/r^2$.

Substituting Eqs.~(\ref{s1}), (\ref{s0u1}) and (\ref{s0v1}) into
Eq.~(\ref{eqncont}) leads to
\begin{equation}
  r^2(\Phi + \Psi)^{''}
  + r\left( {\cal B} + {\cal S} \right) (\Phi + \Psi)^{'}
  - r{\cal S}\Psi^{'}
  + {\cal C} \Psi
  + {\cal D} \Phi = 0,
  \label{eqdiffpsi}
\end{equation}
where:
\begin{equation}
{\cal B} = 1 + {\cal V} - \frac{d\ln\Omega}{d\ln r},
\label{calb}
\end{equation}
\begin {eqnarray}
  \left. {\cal C} \right. 
  &=& -\frac{D}{c_s^2/r^2} -
  2m\frac{\Omega}{\Delta\omega}
  \left( {\cal V} + 2{\cal S} \right) - {\cal B}{\cal S} \nonumber\\
  & & +{\cal S}^2 \left[ (r/{\cal S})^{'} - 1\right] -m^2\left(1+{\cal
      S}{\cal P}\frac{c_s^2 / r^2}{\Delta\omega^2}\right),
    \label{calc}
\end {eqnarray}
\begin{equation} 
  {\cal D} = - 2m\frac{\Omega}{\Delta\omega} \left(
    {\cal V} + {\cal S} \right) - m^2\left(1+{\cal S}{\cal
      P}\frac{c_s^2 / r^2}{\Delta\omega^2}\right),
\label{cald}
\end{equation}
and
\begin{equation}
{\cal V} = \frac{d\ln {\cal F}}{d \ln r}.
\label{calv}
\end{equation}

Eq.~(\ref{eqdiffpsi}) reduces to the equation~(15) of \citet{li00} if
one considers the propagation of free waves ($\Phi=0$), while it
reduces to the equation~(13) of \citet{gt79} in the case of a
homentropic (${\cal S}=0$) flow.

\subsection {Corotation torque}
\label{sec:cortq}

We now estimate the rate of angular momentum exchanged between the
perturber and the radiatively inefficient disk described in
section~\ref{sec:basic}.  This rate therefore corresponds to the disk
torque, which we denote by $\Gamma$, and which we define as the torque
exerted by the disk on the perturber (unless otherwise stated).  It
reads:
\begin{equation}
  \Gamma = \int_{\rm disk} \Sigma_1(r,\varphi) 
  \frac{\partial \Phi}{\partial \varphi} rdrd\varphi.
\label{eqn:torque}
\end{equation}
We limit ourselves to the torque exerted by the disk material lying in
the vicinity of corotation, hence to the corotation torque, which we
denote by $\Gamma_c$. In a linear analysis, this torque can be expressed
as a series of contributions at each azimuthal wavenumber: $\Gamma_c =
\sum_m \Gamma_{c, m}$. Each individual torque can be expressed,
assuming that $\Phi$ is real, as:
\begin{equation}
  \Gamma_{c, m} = m\pi r_c^2 \Phi(r_c) 
  \int_{-\infty}^{\infty} dx\,\Im[\Sigma_1(x)],
  \label{gammam}
\end{equation}
where $\Im$ denotes the imaginary part, $r_c$ is the corotation
radius, and $x=(r-r_c)/r_c$.  We substitute $\Sigma_0u_1$ in
Eq.~(\ref{s1}) by the expression given by Eq.~(\ref{s0u1}), and we
keep only the terms which are large in the vicinity of corotation.  As
in \citet{gt79}, we assume that the disk responds to a slowly
increasing perturbation and take $\Delta\omega$ to have a small,
positive imaginary part $\alpha$:
\begin{equation}
  \Delta\omega = m(\Omega_p-\Omega) +
  i\alpha \approx -mr_c\Omega^{'}(r_c) (x+i\epsilon),
  \label{domega}
\end{equation}
where $\epsilon = -\alpha / [mr_c\Omega'(r_c)] > 0$.  In the vicinity
of corotation, we can finally write:
\begin{equation}
  \Sigma_1(x) = \Psi(x)\left[\frac{\Sigma_0}{c_s^2}\right]_{r_c}
  - \frac{(\Phi+\Psi)(x)}{x +i\epsilon}\left[\frac{2{\cal F}
      {\cal S}}{r^3\Omega^{'}}\right]_{r_c}.
  \label{s1approx}
\end{equation}
We are primarily interested in the imaginary part of $\Sigma_1$. In
the limit $\epsilon\rightarrow 0$, we can write the terms that yield a
non-vanishing contribution to the torque as:
\begin{equation}
  \Im[\Sigma_1(x)] = \Im[\Psi(x)]\left[\frac{\Sigma_0}{c_s^2}\right]_{r_c}
  +\pi\delta(x)\left[\frac{2{\cal F}{\cal S}[\Phi+\Re(\Psi)]}{r^3\Omega^{'}}\right]_{r_c}
  -\frac{\Im[\Psi(x)]}{x}\left[\frac{2{\cal FS}}{r^3\Omega'}\right]_{r_c},
\label{ims1approx}
\end{equation}
where $\delta(x)$ is Dirac's delta function.  The first two terms of
the R.H.S. of Eq.~(\ref{ims1approx}) yield respectively the following
contributions to the corotation torque:
\begin{eqnarray}
  \label{gammam1}
  \Gamma_{c, m, 1}&=& \left[\frac{m\pi\Sigma_0 r^2 \Phi}{c_s^2}\right]_{r_c} \int_{-\infty}^{\infty} dx\,\Im[\Psi(x)]\\
  \label{gammam2}
  \Gamma_{c, m, 2}&=& \left[\frac{2m\pi^2{\cal F}{\cal S}\Phi(\Phi+\Re(\Psi))}{r \Omega'}\right]_{r_c}.
\end{eqnarray}

The third term of Eq.~(\ref{ims1approx}) yields a contribution that
can be shown to be negligible, in the planetary context, compared to
$\Gamma_{c,m,2}$. This is shown in appendix~\ref{apA}.

The first term, $\Gamma_{c, m, 1}$, is the contribution of the
function $\Psi$, such as in the barotropic case.  The second term,
$\Gamma_{c, m, 2}$, corresponds to a singularity at corotation,
associated to a non-vanishing entropy gradient. It corresponds to the
torque arising from the advection of entropy in the corotation region,
which results in a surface density perturbation if the entropy is not
uniform. The perturbation is singular for the surface density and the
entropy, but not for the pressure (see section~\ref{sec:gen}). It
remains confined to corotation, where it yields a singular
contribution to the torque. Some further insight into the dynamics of
this perturbation will be given in section~\ref{sec:dyncor}.

We provide in the next section an expression for the corotation torque
in the limit of a cold disk, then we turn to the general case.

\subsubsection{Limit of a cold disk}
\label{sec:cold}

We contemplate here the case for which $|\Psi| \ll |\Phi|$, which we
shall refer to as a cold case. This condition depends on the strength
of the perturbing potential, its radial scale, and on the disk
temperature.  In particular, in the planetary context, some corotation
resonances may correspond to a cold situation, while others have
$|\Psi|\sim|\Phi|$. Nevertheless, a given resonance eventually
satisfies the cold case condition as the disk temperature tends to
zero.

The evaluation of Eq.~(\ref{gammam1}) requires an explicit
expression for $\Psi$, obtained by solving the differential equation
(\ref{eqdiffpsi}) in the vicinity of corotation.  This has been done
by \citet{gt79} for a cold barotropic disk.  An explicit solution can
also be obtained for a cold adiabatic disk within the same level of
approximation.  Note however that some additional difficulties arise,
in particular the existence of a double pole (term proportional to
$\Delta\omega^{-2}$) in the coefficients ${\cal C}$ and ${\cal D}$,
defined respectively by Eqs.~(\ref{calc}) and~(\ref{cald}).

We discard the double pole for the following reasons:
\begin{itemize}
\item Unlike the simple pole, it scales with $c_s^2$, which indicates
  that when the disk aspect ratio tends to zero, it becomes
  negligible; differently stated, there should be a critical disk
  thickness under which it is safe to neglect this term.
\item This term is the only one that depends both on the entropy and
  on the pressure gradients. As we shall see in
  section~\ref{sec:sgrad}, our results of numerical simulations for a
  planet embedded in a disk with aspect ratio $h=0.05$ show that the
  torque excess with respect to an isothermal situation essentially
  depends on ${\cal S}$, the gradient of entropy, which indicates that
  already for $h=0.05$ the double pole term is negligible.
\item The double pole is regularized with a very small amount of
  dissipation. Even the molecular viscosity suffices to render it
  negligible in the disks that we consider (S.-J. Paardekooper, private
  communication).
\end{itemize}
Discarding the double pole, and within the same level of approximation
as \citet{gt79}, Eq.~(\ref{eqdiffpsi}) can be recast, in the vicinity
of the corotation, as
\begin{equation}
  \frac{d^2\Psi}{dx^2} - q^2\Psi = -\frac{P_1 \Phi(r_c)}{x+i\epsilon},
\label{eqdiffpsirc}
\end{equation}
where
$$
P_1 = \left[ \frac{2\Omega}{r\Omega^{'}} \left( {\cal V} + {\cal S}
  \right) \right]_{r_c}{\rm and~} q = (Dr/c_s)_{r_c} \approx (\kappa
r/c_s)_{r_c}.
$$
The general solution of Eq.~(\ref{eqdiffpsirc}) reads
\begin{eqnarray}
  \left. \Psi(x)
  \right. &=& \frac{P_1}{2q}\Phi(r_c) \left[ e^{qx}\int_x^{\infty} \frac{dt}{t+i\epsilon}e^{-qt}  \right. \nonumber\\
  & & \left. + e^{-qx}\int_{-\infty}^{x} \frac{dt}{t+i\epsilon}e^{qt}\right],
\label{psicold}
\end{eqnarray}
which reduces to the equation~(53) of \citet{gt79} when ${\cal S}=0$.
Combining Eqs.~(\ref{gammam1}) and (\ref{psicold}) yields the
contribution $\Gamma_{c,m,1}$ to the corotation torque:
\begin{equation}
\Gamma_{c,m,1} = \Gamma_0 \left[ ({\cal V}+{\cal S})\,\Phi^2 \right]_{r_c},
\label{gcm1cold}
\end{equation}
where $\Gamma_0 = -(m \pi^2 \Sigma_0) / (2 B r\Omega^{'})$ is to be
evaluated at the corotation radius.  It can be approximated as
$(4m\pi^2\Sigma_0/3\Omega^2)_{r_c}$ in a Keplerian disk.

The second contribution to the corotation torque, given by
Eq.~(\ref{gammam2}), is specific to the adiabatic case and involves
the singularity arising from the entropy advection. Using
Eq.~(\ref{gammam2}) and noting that $|\Re(\Psi)| \ll |\Phi|$, this
contribution to the corotation torque reads
\begin{equation}
  \Gamma_{c,m,2} = -\Gamma_0 \left[ {\cal S}\,\Phi^2 \right]_{r_c}.
\label{gcm2cold}
\end{equation}
From Eqs.~(\ref{gcm1cold}) and (\ref{gcm2cold}), we infer the
corotation torque for a cold, adiabatic disk, which reads:
\begin{equation}
\Gamma_{c,m} = \Gamma_0 \left[ {\cal V}\,\Phi^2 \right]_{r_c}.
\label{gcmcold}
\end{equation}
This expression does not depend on ${\cal S}$.  We note from
Eqs.~(\ref{calf}) and~(\ref{calv}) that ${\cal V}$ can be approximated
as
\begin{equation}
{\cal V} = \frac{d\ln \Sigma_0 / B}{d\ln r},
\end{equation}
since the disk aspect ratio at corotation $h(r_c) = c_s(r_c) / [r_c
\Omega(r_c)]$ satisfies $h(r_c) \ll 1$.  Eq.~(\ref{gcmcold}) therefore
corresponds to the corotation torque expression\footnote{They have a
  negative sign because they consider the torque exerted by the
  perturber on the disk.}  of \citet{gt79}.  This argues that the
corotation torque for a cold case does not depend on whether the disk
can radiate energy efficiently (assuming a locally isothermal equation
of state) or not (assuming an adiabatic energy equation).  This can be
expected on general grounds: in the cold disk limit, the internal
energy of the fluid is negligible with respect to its mechanical
energy, hence the power (and the torque) of the tidal force correspond
to the case of non-interacting test particles, for which the
expression of \citet{gt79} prevails.

\subsubsection {General case}
\label{sec:gen}

We consider in this section the general case where we cannot neglect
$\Psi$ with respect to $\Phi$ in Eqs.~(\ref{gammam1})
and~(\ref{gammam2}), as we have done in the previous section.  Instead
of resorting to a solution of Eq.~(\ref{eqdiffpsi}), we shall use a
method similar to the method used by \citet{tanaka}, based on the jump
of angular momentum flux at corotation.  In the case of \cite{tanaka},
this eventually yields a torque expression similar to the expression
of \citet{gt79}, except that $\Phi$ has to be substituted by
$\Phi+\eta$ (where $\eta$ is the enthalpy perturbation). The drawback
of this method is that it provides a torque expression that depends on
the (unknown) solution of the differential equation. Nevertheless, it
gives some insight into the dynamics of the corotation region, and
allows to draw the general trends of the corotation torque in an
adiabatic disk. In our case, the torque expression features
$\Psi=p_1/\Sigma_0$. We note that in the isothermal case, \citet{zl06}
have provided an explicit solution for the perturbed enthalpy at
corotation, that leads to a corotation torque expression that only
depends on the forcing potential.

We note that the jump of angular momentum flux at corotation misses
the singular contribution of the entropy perturbation at corotation
and as such leads only to an evaluation of $\Gamma_{c,m,1}$.  The
contribution $\Gamma_{c,m,2}$ of the entropy perturbation to the
corotation torque needs to be calculated similarly as in
Eq.~(\ref{gcm2cold}).  The angular momentum flux $F_A$ is given by:
\begin{equation}
  F_A = \Sigma_0 r^2 \int_0^{2\pi} \Re(u)\Re(v)d\varphi = \pi\Sigma_0 r^2 \Re(uv^*),
\label{fa}
\end {equation}
where $\Re$ stands for the real part and the star superscript denotes
the complex conjugate.  Eq.~(\ref{fa}) can be written as $F_A = \sum_m
F_{A,m}$ with:
\begin{equation}
  F_{A,m} = \pi \Sigma_0 r^2 \left[ \Re(u_{1})\Re(v_{1}) +
  \Im(u_{1})\Im(v_{1}) \right].
\label{fam}
\end{equation}
Combining Eqs.~(\ref{s0u1}), (\ref{s0v1}) and~(\ref{fam}), we obtain
\begin{eqnarray}
  \left. F_{A,m}
  \right. &=& \frac{m\pi\Sigma_0 r}{D}\left[
    \Im(\Phi+\Psi)\frac{d\Re(\Phi+\Psi)}{dr} \right. \nonumber\\ & -
  &\left. \Re(\Phi+\Psi)\frac{d\Im(\Phi+\Psi)}{dr} \right. \nonumber\\ &
  + & \left. \frac{{\cal S}}{r} \left\{ \Re(\Phi)\Im(\Psi) -
      \Im(\Phi)\Re(\Psi) \right\} \right].
\label{fam2}
\end{eqnarray}
In the homentropic (${\cal S}=0$) case, Eq.~(\ref{fam2}) reduces to the 
expression used by \citet{tanaka}.  The contribution $\Gamma_{c,m,1}$ 
to the corotation torque is then given by:
\begin{equation}
  \Gamma_{c,m,1} = \lim_{r_c^+,r_c^-\rightarrow r_c}
  [F_{A,m} (r_c^+) - F_{A,m} (r_c^-)],
\label{gc1}
\end{equation}
where $r_c^+>r_c$ and $r_c^-<r_c$ are the radii of locations
respectively beyond and before corotation, and where we evaluate
the flux of advected angular momentum.

\citet{tanaka} showed that $\Phi+\eta$ is continuous at corotation.
Here, since Eq.~(\ref{eqdiffpsi}) cannot be recast as an ordinary
differential equation involving only $\Phi+\Psi$, we have to consider
more stringent albeit reasonable assumptions, namely that both $\Phi$
and $\Psi$ are continuous at corotation. The fact that $\Phi$ is
continuous at corotation can be realized with an arbitrarily small
softening length of the potential, in the case of an embedded
point-like mass (for which the potential components would diverge
logarithmically at corotation, in the absence of any
softening). Assuming that $\Phi$ is continuous at corotation,
Eq.~(\ref{eqdiffpsi}) imposes that $\Psi$ is also continuous at
corotation (we would otherwise have a null linear combination of
$\delta(x)$ and $\delta'(x)$ functions with non-vanishing
coefficients, which is impossible).

Our continuity assumption implies that the terms proportional to
${\cal S}$ in the R.H.S. of Eq.~(\ref{fam2}) does not contribute to
the torque.  The jump in the advected flux therefore comes from the
jump in $d(\Phi+\Psi)/dr$.

We integrate Eq.~(\ref{eqdiffpsi}) over an infinitesimal interval
containing $r=r_c$. All finite terms in this equation yield a vanishing
contribution, hence we are left only with the jump of
$d(\Phi+\Psi)/dr$ stemming from the second derivative term of
Eq.~(\ref{eqdiffpsi}) and the poles of the terms ${\cal C}\Psi$ and
${\cal D}\Phi$. This reads:
\begin{equation}
  \frac{d(\Phi+\Psi)}{dr}(r_c^+) - \frac{d(\Phi+\Psi)}{dr}(r_c^-) = \frac{i\pi}{r_c}\left[ P_2(\Phi+\Psi)(r_c) - Q\Phi(r_c) \right],
\label{jump}
\end{equation}
where
$$ P_2 = \left[ \frac{2\Omega}{r\Omega^{'}} \left( {\cal V} +
    2{\cal S} \right) \right]_{r_c} \rm{~and~~} Q = \left[
  \frac{2\Omega}{r\Omega^{'}}{\cal S} \right]_{r_c}.
$$
Using Eqs.~(\ref{fam2}), (\ref{gc1}), (\ref{jump}) and
$2B=\kappa^2/2\Omega$, we find that
\begin{equation}
  \Gamma_{c,m,1} = \Gamma_0 \left[ \left\{ {\cal V} +
      2{\cal S} \right\} |\Phi+\Psi|^2 - {\cal S}\,\Phi\,\Re(\Phi+\Psi)
  \right]_{r_c}.
\label{gcm1}
\end{equation}
Eq.~(\ref{gcm1}) reduces to Eq.~(\ref{gcm1cold}) in the cold disk limit.

We now come to the contribution $\Gamma_{c,m,2}$ of the entropy
perturbation to the corotation torque. Eq.~(\ref{gammam2})
yields:
\begin{equation}
  \Gamma_{c,m,2}
  =-\Gamma_0\left[{\cal S}\,\Phi\,\Re(\Phi+\Psi)\right]_{r_c}.
\label{gcm2}
\end{equation}
Eq.~(\ref{gcm2}) reduces to Eq.~(\ref{gcm2cold}) in the cold disk limit.

The general expression for the corotation torque is obtained by
accounting for the contribution given by Eq.~(\ref{gcm1}), and that of the
entropy perturbation, given by Eq.~(\ref{gcm2}):
\begin{equation}
  \Gamma_{c,m} = \Gamma_0 \left[ \left\{
      {\cal V} + 2{\cal S} \right\} |\Phi+\Psi|^2 -
    2{\cal S}\,\Phi\,\Re(\Phi+\Psi) \right]_{r_c}.
\label{gcm}
\end{equation}
Eq.~(\ref{gcm}) reduces to the expression of \citet{tanaka} when
${\cal S}=0$, while it reduces to that of \citet{gt79} for a cold
disk.

A case of interest is that of a disk perturbed by a peaked potential
(that of an embedded protoplanet for instance), for which $|\Phi+\Re(\Psi)|
\ll |\Phi|$, and  $|\Phi+\Re(\Psi)| \ll |\Re(\Psi)|$ at corotation. 
 For such case, $|\Phi\Re(\Phi+\Psi)|_{r_c} \gg |\Phi+\Psi|_{r_c}^2$, 
 hence the corotation torque may be approximated as $\Gamma_{c,m} \approx -
2\Gamma_0[{\cal S}\,\Phi\,\Re(\Phi+\Psi)]_{r_c}$.  The corotation torque may
therefore be much larger in the non-homentropic case (${\cal S}\ne
0$) than in the homentropic case (${\cal S}=0$).  Furthermore, its
sign is given by that of ${\cal S}$ rather than that of
${\cal V}$.  This enhancement of the corotation torque in an
adiabatic flow may have a dramatic impact on the type~I migration of
an embedded protoplanet, as was noted by PM06.

\section {Numerical study of an isolated corotation resonance}
\label{sec:numiso}

We check in this section the analytical predictions of section
\ref{sec:isores} by means of numerical simulations involving an
isolated corotation resonance (hereafter CR).

\subsection {Numerical issues}
\label{sec:numissues}

Our setup offers a number of similarities with the setup of
\citet{mo04} for the case of an isothermal disk.  The hydrodynamics
equations for the disk described in section \ref{sec:basic} are solved
using the code {\sc Fargo}. A description of the properties of this
code is deferred to section~\ref{sec:plasetup}, in which the code is
used to simulate an embedded planet. As in \citet{mo04}, we deal with
the $m=3$ CR.  The disk is therefore torqued by an $m=3$ external
potential $\Phi$ that reads
\begin{equation}
  \Phi(r,\varphi,t) = T(t/\tau)\phi(r)\cos[3(\varphi-\Omega_p t)],
\end{equation}
where $\phi(r)$ denotes the radial profile of the potential,
$\Omega_p$ its pattern speed (note that we work in the corotating
frame), $t$ is the time and where
\begin{eqnarray*}
T(x)  & = & \sin^2(\pi x/2) \mbox{~if~} x < 1\\
      & = & 1 \mbox{~otherwise}
\end{eqnarray*}
is a temporal tapering that turns on the potential on the timescale
$\tau$.

The total torque $\Gamma_c$ exerted by the disk on the
perturber, given by Eq.~(\ref{eqn:torque}), is evaluated by
\begin{equation}
  \Gamma_c = \sum_{i=0}^{N_r-1} \sum_{j=0}^{N_s-1}
  \frac{\Phi_{i,j+1}-\Phi_{i,j-1}}{2\Delta\varphi} \Sigma_{i,j} S_{i,j},
\label{num:torque}
\end{equation}
where $N_r$ ($N_s$) is the radial (azimuthal) number of zones of the
mesh, $S_{i,j}$ is the surface area of zone $(i,j)$, $\Phi_{i,j}$ and
$\Sigma_{i,j}$ are the external potential and surface density at the
center of this zone, and $\Delta\varphi = 2\pi / N_s$ is the azimuthal
resolution.  Furthermore, the contribution $\Gamma_{c,1}$ of
the function $\Psi$ to the torque is obtained by substituting $\Sigma_1$
by $p_1/c_s^2$ in Eq. (\ref{eqn:torque}). It is therefore evaluated by
\begin{equation}
  \Gamma_{c,1} = \sum_{i=0}^{N_r-1} \sum_{j=0}^{N_s-1}
  \frac{\Phi_{i,j+1}-\Phi_{i,j-1}}{2\Delta\varphi}
  \frac{p_{i,j}}{c_{si,j}^2} S_{i,j},
\label{num:torquepsi}
\end{equation}
where $p_{i,j}$ and $c_{si,j}$ are the pressure and sound speed at the
center of zone $(i,j)$.  The contribution $\Gamma_{c,2}$ of the
entropy perturbation to the torque is eventually estimated as follows:
\begin{equation}
  \Gamma_{c,2} = \Gamma_c - \Gamma_{c,1}.
  \label{num:torquedc}
\end{equation}

The radial computational domain is narrow enough to avoid the location
of the $m=3$ inner and outer Lindblad resonances \citep[see][]{mo04}.
Despite this precaution, wave killing zones next to the boundaries
were implemented to minimize unphysical wave
reflections \citep{valbo06}. Furthermore, the torque evaluation is performed
by summing only on a domain of the grid that does not contain the
wave killing zones, and the summation includes a spatial tapering on
the edges of that domain.

The disk surface density and temperature are initially axisymmetric
with power-law profiles: 
\begin{equation}
\label{eqn:siglaw}
\Sigma (r) = \Sigma_c \,(r / r_c)^{-\sigma}
\end{equation}
and 
\begin{equation}
\label{eqn:tlaw}
T(r) = T_c \,(r / r_c)^{-1+2f},
\end{equation}
where $\Sigma_c$ and $T_c$ are the surface density and temperature at
the corotation radius $r_c$, and where $f$ is the flaring index of the
disk. The disk aspect ratio is given by $h(r) = H(r) / r = h(r_c)
(r/r_c)^{f}$, where $H(r)$ is the disk scale height at radius $r$. 
A vanishing value of the flaring index $f$ therefore 
corresponds to a uniform disk aspect ratio. The functions
${\cal V}$ and ${\cal S}$ are constant and read:
\begin{eqnarray}
 {\cal V}&=&3/2-\sigma\label{eqV}\\
 {\cal S}&=&\sigma-(\sigma+1-2f)/\gamma\label{eqS}.
\end{eqnarray}

The main numerical parameters are those taken by \citet{mo04}, namely
a $h(r_c) = 0.01$ disk aspect ratio at corotation, and $\Sigma_c = 1$.
Our disk is inviscid.  The libration islands are resolved by $30$
zones azimuthally.  As the potential increases, the radial width of
the islands also increases.  Their maximal radial width $W$ spans
approximately $20$ zones.

The results presented in next section have the following units: the
mass of the central object $M_*$ is the mass unit, the
corotation radius $r_c$ of our $m=3$ CR is the distance unit and the
Keplerian orbital period $T_{\rm orb}$ at $r=r_c$ is $2\pi$ times the time unit.

\subsection {Results}
\label{sec:resiso}

\begin{figure}
  \plotone{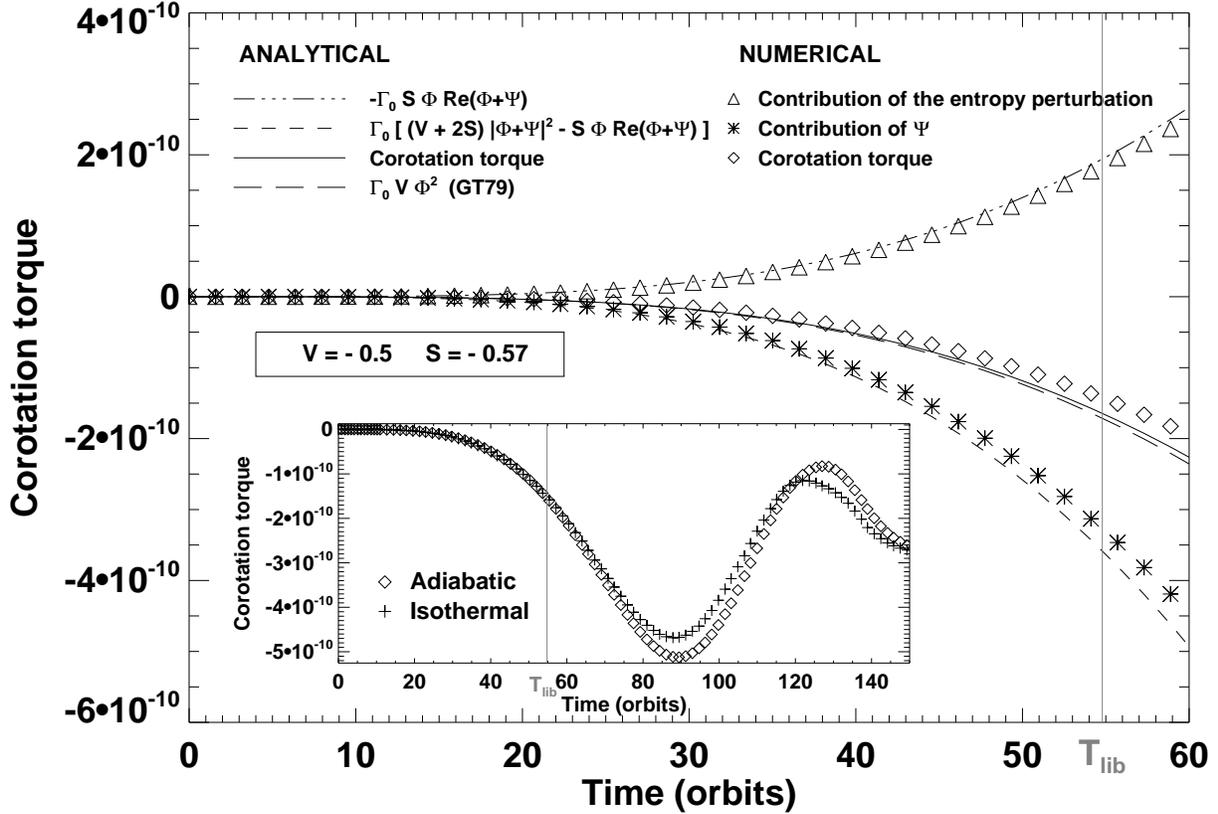} \figcaption{\label{localflat} Corotation torque
    exerted by the disk on the perturber, as a function of time,
    assuming a flat radial profile of the potential.  The results
    shown are obtained with an adiabatic calculation, except in the
    close-up, where we compare the isothermal and adiabatic corotation
    torques over the whole duration of the calculations.  Numerical
    results are displayed with a symbol while the theoretical
    expectations are displayed with curves.  We plot as a function of
    time the adiabatic corotation torque (diamonds and solid curve),
    the contribution of the function $\Psi$ to the torque (stars and
    dashed curve), and the contribution of the entropy perturbation
    (triangles and dot-dashed curve).  The long-dashed curve, which is
    nearly superimposed to the solid curve, displays the corotation
    torque expression of \citet{gt79}.  The vertical solid line gives
    an estimate of the final libration time (see text).}
\end{figure}
\begin{figure}
  \plottwo{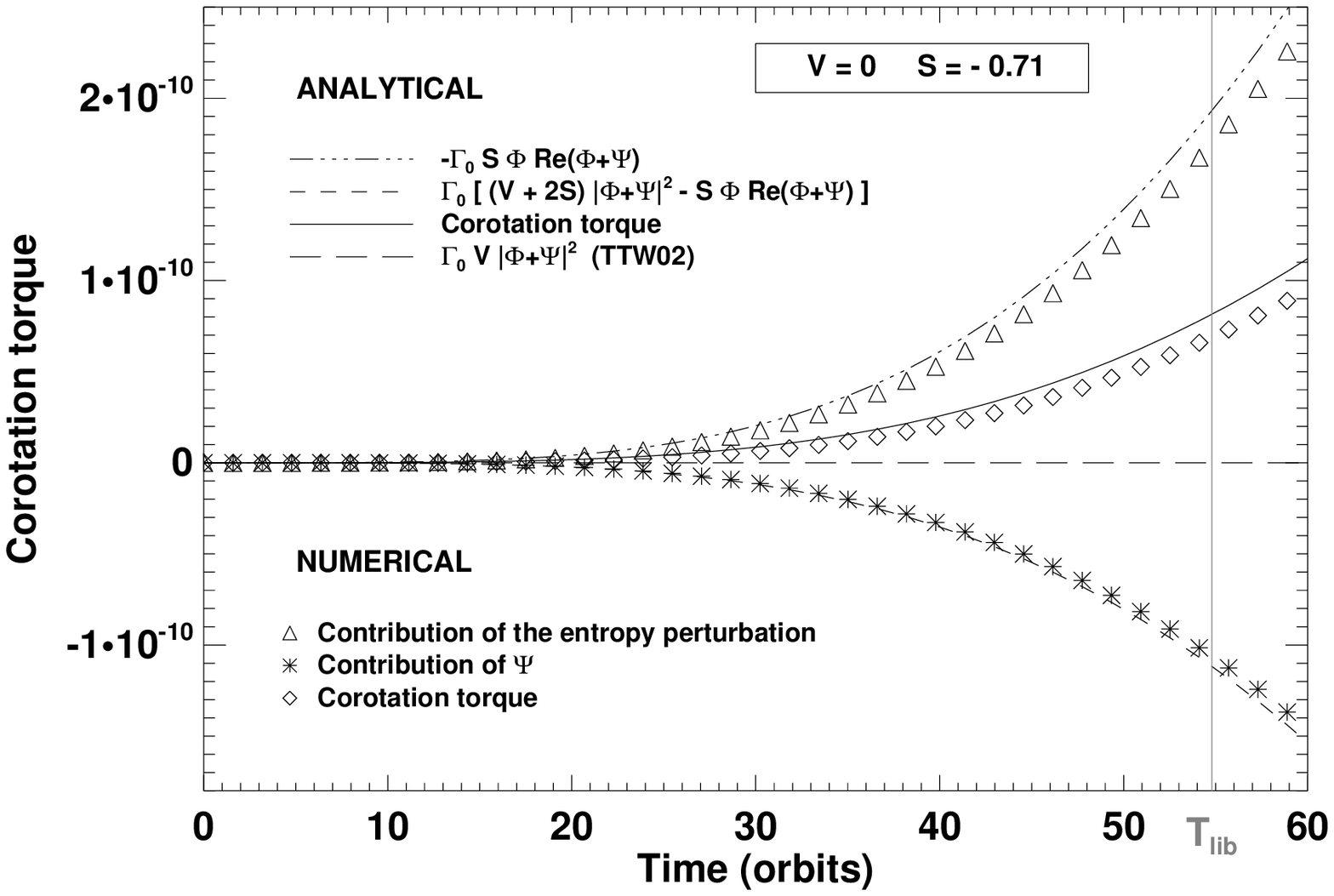}{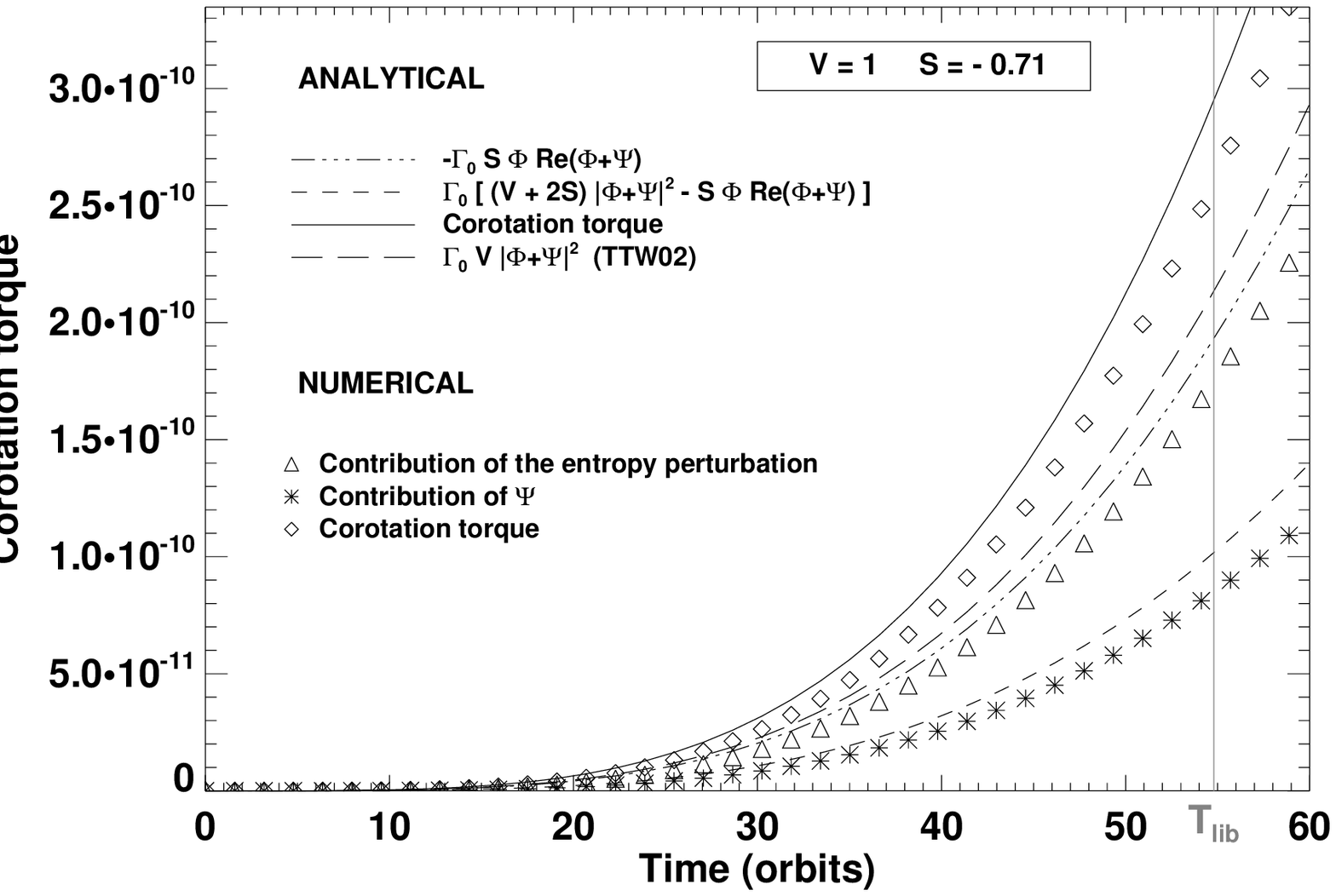} \figcaption{\label{localnoflat} Same as
    Fig.~\ref{localflat}, except that the results are obtained with a
    peaked potential, with ${\cal V}=0$ ( ${\cal V}=1$) in the left
    (right) panel.  The long-dashed curve in both panels shows the
    expectation from the corotation torque expression of
    \citet{tanaka}, denoted by TTW02.}
\end{figure}

We consider three cases, corresponding respectively to Figs.~\ref{localflat}, 
\ref{localnoflat}a and~\ref{localnoflat}b:
\begin{enumerate}
\item An external potential with flat profile $\phi(r) = -10^{-5}$, as
  in \citet{mo04}.  This case, that we call the ``flat potential
  case'', has the following parameters: $\sigma=2$ and $f=-0.3$, which
  implies, from Eqs.~(\ref{eqV}) and~(\ref{eqS}), that ${\cal V} =
  -0.5$ and ${\cal S} \approx -0.57$,
\item A potential profile that corresponds to the $m=3$ Fourier
  component of the smoothed potential of a $M = 3.1\times10^{-6} M_*$
  point-like object.  The softening length is $\varepsilon = H(r_c)$,
  which is approximately equal to $W$. The object rotates at speed
  $\Omega_p$, with orbital radius $r_c$. This neglects the pressure
  gradient effects, as we do not resolve the distance from orbit to
  corotation, but this distance is much smaller than the potential
  softening length, so this is not a concern in the present case.  By
  contrast to the previous case, we call this situation the ``peaked
  potential case''.  The value of $M$ was chosen so that $\phi(r_c) =
  -10^{-5}$, as in the flat potential case.  For this calculation we
  have $\sigma=1.5$ and $f=-0.3$, so that ${\cal V} = 0$ and ${\cal S}
  \approx -0.71$.  The results are depicted in
  Fig.~\ref{localnoflat}a.
\item A calculation similar to the previous one, except that
  $\sigma=0.5$ and $f=-0.1$, so that ${\cal V} = 1$ and ${\cal S}
  \approx -0.71$. The results are depicted in Fig.~\ref{localnoflat}b.
\end{enumerate}

For the three pairs $({\cal V},{\cal S})$ quoted above, the
tapering timescale value is $\tau = 150\,T_{\rm orb}$, 
which corresponds to the duration of the calculations.  
This is about three times larger than the final libration time, estimated as
\begin{equation}
  T_{\rm lib} \sim \frac{1}{m} \left( \frac{3|\phi(r_c)|}{32} \right)^{-1/2} \approx 55\,T_{\rm orb}.
  \label{tlib}
\end{equation}
In each case we evaluate:
\begin{itemize}
\item the total corotation torque (diamonds) with
  Eq.~(\ref{num:torque}), to be compared to the analytical expression
  (solid curve) given by Eq.~(\ref{gcm}). In our units, $\Gamma_0
  \approx 39.5$,
\item the contribution of the function $\Psi$ to the corotation torque
  (stars) obtained with Eq.~(\ref{num:torquepsi}), the expected
  expression of which (dashed curve) is calculated using
  Eq.~(\ref{gcm1}),
\item the contribution of the entropy perturbation to the torque
  (triangles) using Eq.~(\ref{num:torquedc}), which is to compare to
  the prediction of Eq.~(\ref{gcm2}), represented by the dot-dashed
  curve.
\end{itemize}

In these figures, the corotation torque first increases with time
since the potential is progressively turned on until it reaches its
final value at the end of the calculation.  After some time it starts
to oscillate.  This oscillation corresponds to the saturation of the
CR, as the ratio $t/T_{\rm lib}$ tends to unity \citep{ol03}.
Figs. \ref{localflat}, \ref{localnoflat}a and~\ref{localnoflat}b
therefore argue that our numerical simulations succeed in reproducing
the results of our analytical study as long as $t \lesssim T_{\rm
  lib}$, that is when a linear analysis is grounded (which requires that
the time of the calculation be much smaller than the libration time).

The examination of the results of these calculations leads to the
following comments:

\begin{itemize}
\item In the flat potential case, depicted in Fig.~\ref{localflat}, we
  have $\Re[\psi(r_c)] \approx -0.02\phi(r_c)$ throughout the
  calculation, where $\psi(r)$ denotes the radial profile of $\Psi$.
  This situation therefore corresponds to a cold case.  As expected
  from Eq.~(\ref{gcmcold}), the analytical corotation torque and the
  expression of \citet{gt79} almost coincide.  The close-up shows the
  torque evolution over the whole extent of the calculation, up to
  $t=\tau$.  The torque obtained with a locally isothermal equation of
  state is also depicted. Our isothermal runs have same radial
  temperature dependence as the adiabatic runs (see
  Eq.~(\ref{eqn:tlaw}). Although there is an entropy gradient in these
  isothermal calculations, it does not contribute to the corotation
  torque as it would in an adiabatic disk: the appearance of the
  singular contribution at corotation in the adiabatic case is linked
  (i) to the advection of entropy and (ii) to the appearance of a
  singularity in the perturbed density and temperature fields.  In the
  isothermal situation, neither the entropy is conserved along a fluid
  element path, nor is a temperature singularity allowed to appear.
  The comparison of isothermal and adiabatic calculations shows that,
  as expected for a cold case, the adiabatic and isothermal torques
  coincide, as long as we are in the linear regime.  We note that both
  torques do not oscillate about $0$ since the potential reaches a
  stationary value only at the end of the calculation.

\item For the two calculations of the peaked potential case, depicted
  in Figs.~\ref{localnoflat}a and~\ref{localnoflat}b, we find that
  $\Re[\psi(r_c)] \approx -0.2\phi(r_c)$. Thus, the term
  $|-\Phi\Re(\Phi+\Psi)|$ slightly dominates the term $|\Phi+\Psi|^2$
  in Eq. (\ref{gcm}).  Because ${\cal S}<0$ for these calculations,
  the corotation torque in the adiabatic case (diamonds and solid
  curve) is larger than the corotation torque in an isothermal disk
  (long dashed curve) with the same parameters, as predicted by
  \citet{tanaka}.  In particular, in the case for which ${\cal V}=0$,
  the isothermal corotation torque vanishes, while we find a net,
  positive corotation torque for an adiabatic flow, in correct
  agreement with the analytical expression.
\end{itemize}

\subsection{Dynamics of the corotation region}
\label{sec:dyncor}

We discuss in this section the dynamics of the corotation resonance of
an adiabatic disk and give some comments about the corotation torque
expression of Eq.~(\ref{gcm}). 

In the isothermal case, the corotation torque expression involves the
product of the gradient of vortensity and the square of the effective 
potential ($\Phi+\eta$), see e.g. \citet{tanaka}. The torque
is then given by the angular momentum budget between material flowing
outwards and material flowing inwards at corotation, regardless of the
sign of $\Phi+\eta$. Eq.~(\ref{gcm}) displays a term that has a
similar behavior, except that it does not feature the vortensity
gradient only, but rather ${\cal V}+2{\cal S}$.  This factor scales
with the (logarithmic) gradient of $(\Sigma_0/B)S^{2/\gamma}$, which
is a key quantity considered by \citet{li00} and by \citet{lovelace},
who pointed out that vortensity is not conserved in a two-dimensional 
adiabatic flow.

\begin{figure}
  \plottwo{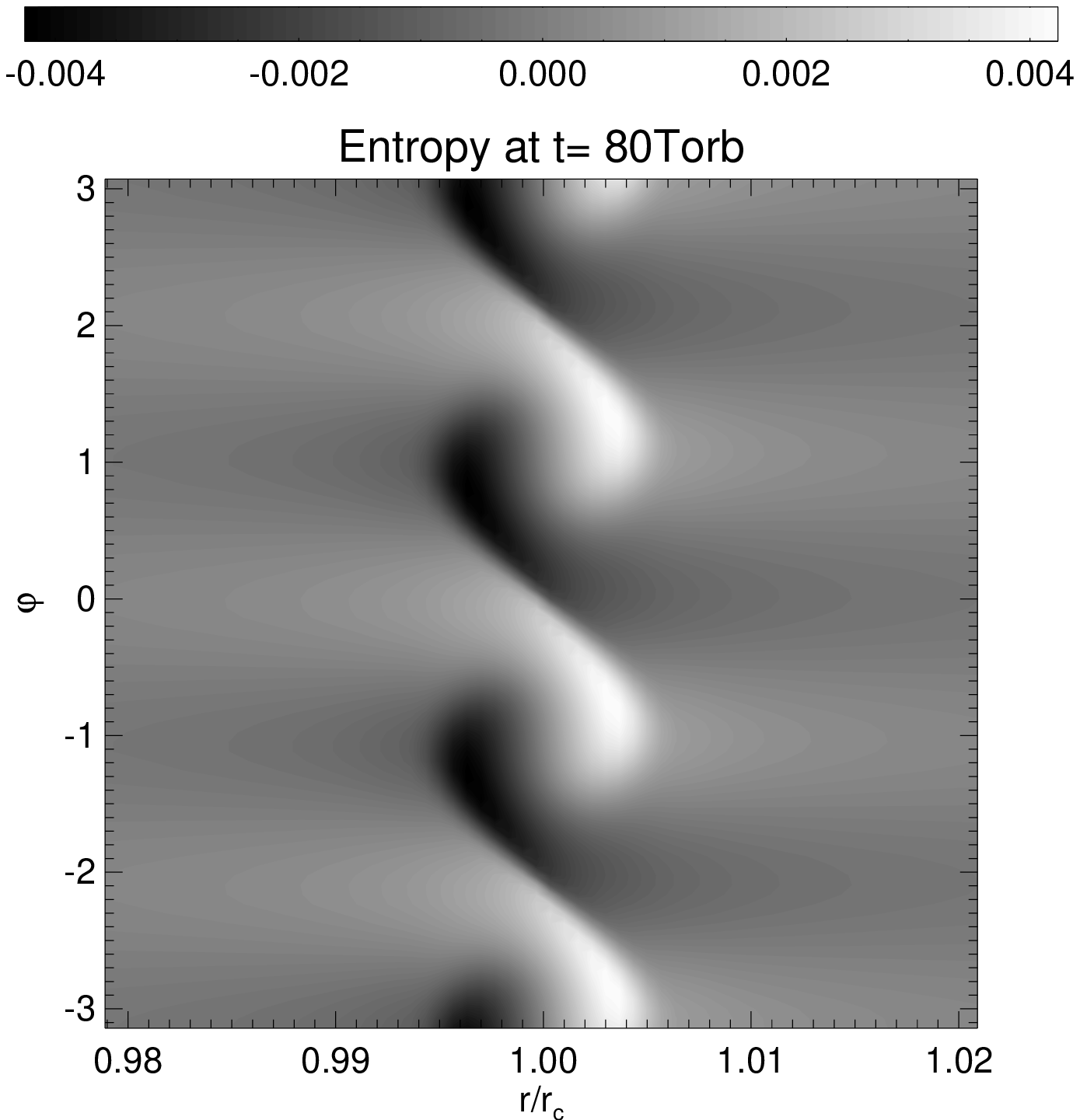}{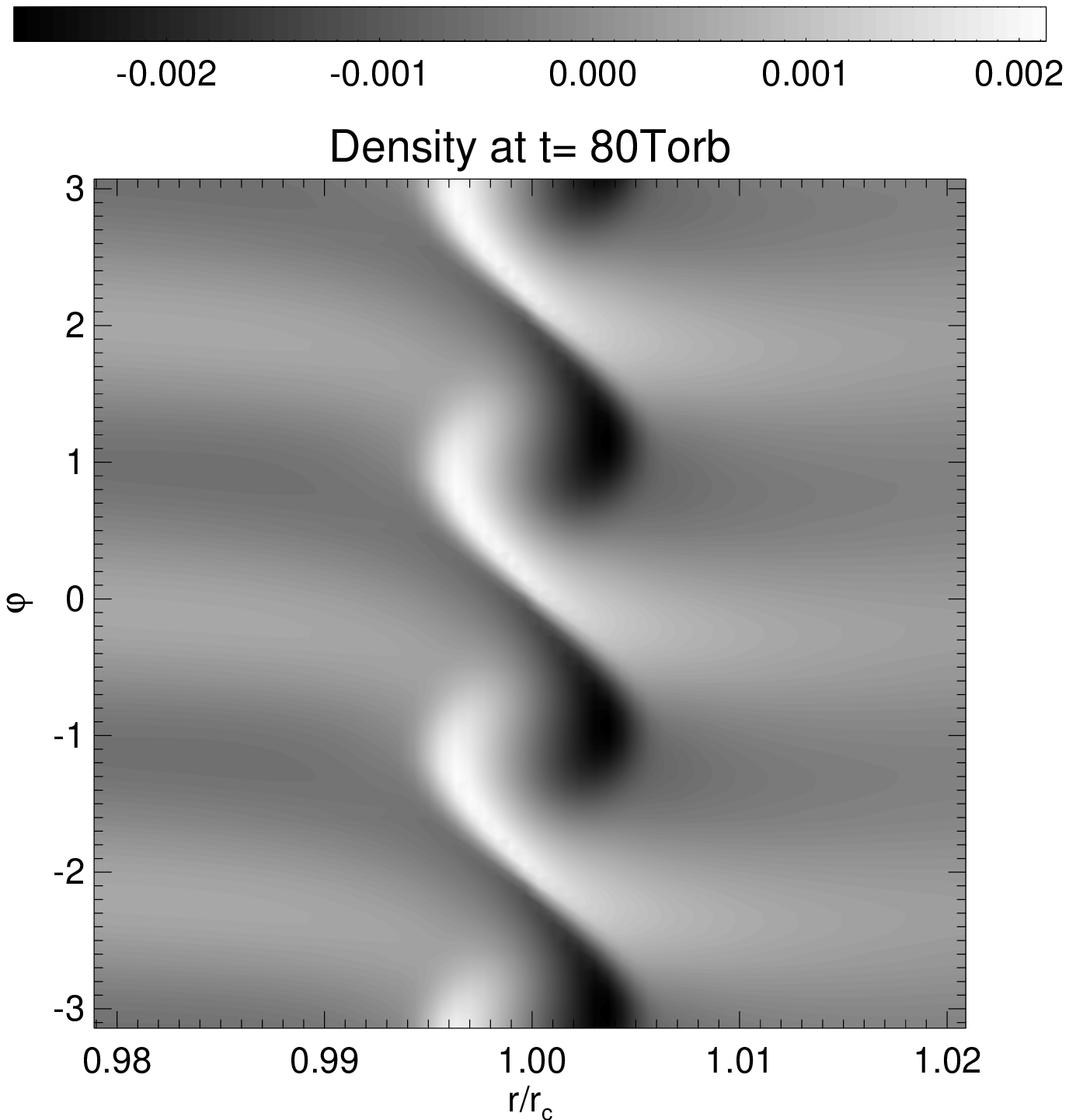} \figcaption{\label{fig:entro}Relative
    perturbation of entropy (left) and surface density (right) for an
    isolated resonance, at $t \approx 1.5\,T_{\rm lib}$. Libration is
    clockwise.}
\end{figure}

In addition to this term, Eq.~(\ref{gcm}) contains a term that scales
with $\Phi[\Phi+\Re(\Psi)]$. The sign of this term therefore depends
on the relative signs of $\Phi$ and $\Phi+\Re(\Psi)$. In order to get
some insight into the physical meaning of this term, we show at
Fig.~\ref{fig:entro} the response of the disk in the corotation
region, for the entropy and the surface density. These fields
correspond to the calculation with the flat potential profile considered at the
previous section.  The disk has a negative radial entropy gradient.
Therefore, libration brings the (larger) inner entropy to the outer
part of the libration islands, yielding a positive perturbed entropy
(brighter areas), while it brings the (smaller) outer entropy to the
inner part of the libration islands, yielding a negative perturbed
entropy (darker areas). An opposite behavior is observed for the
perturbed density, since the relative pressure perturbation (not
represented) is much smaller.

\begin{figure}
  \plottwo{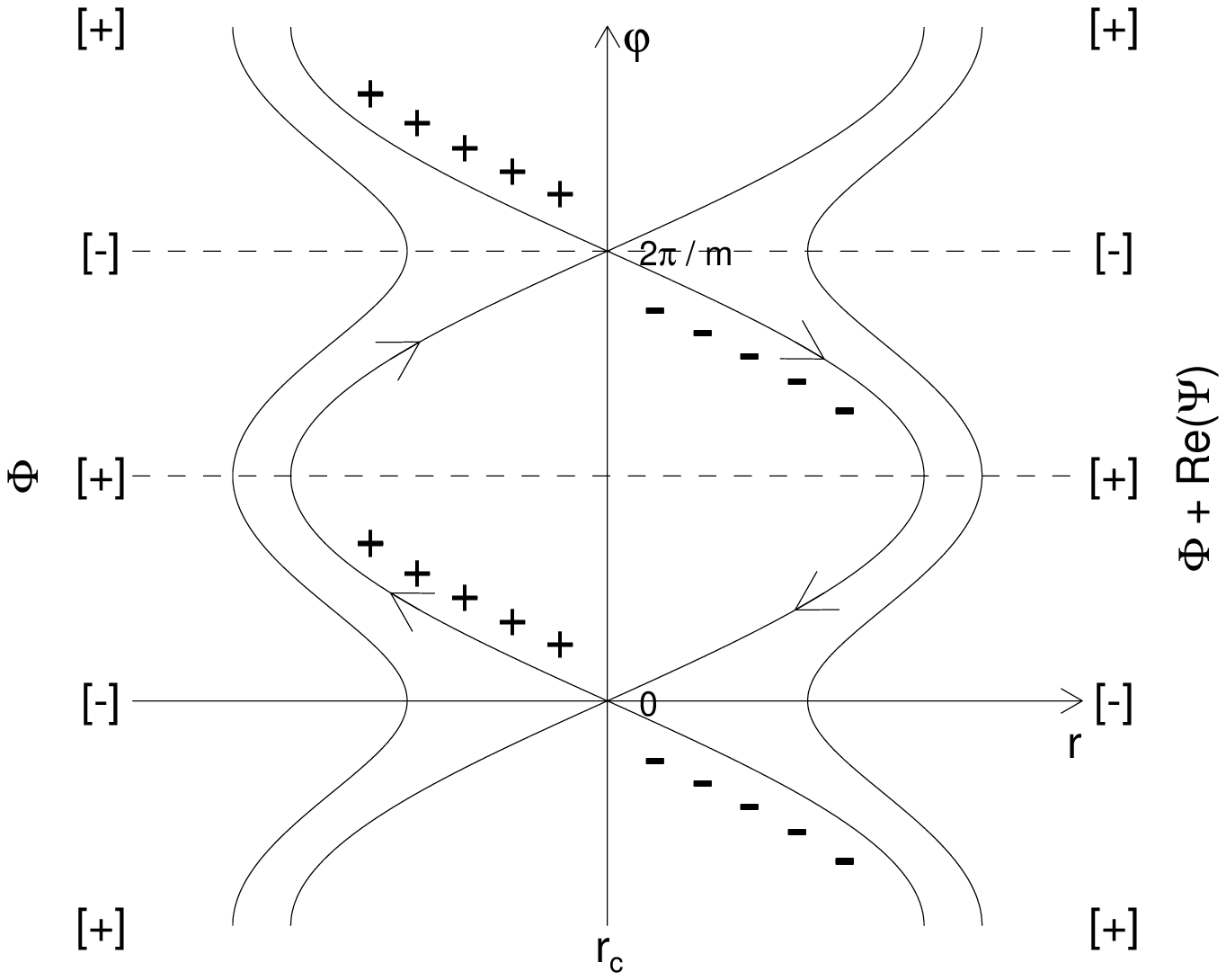}{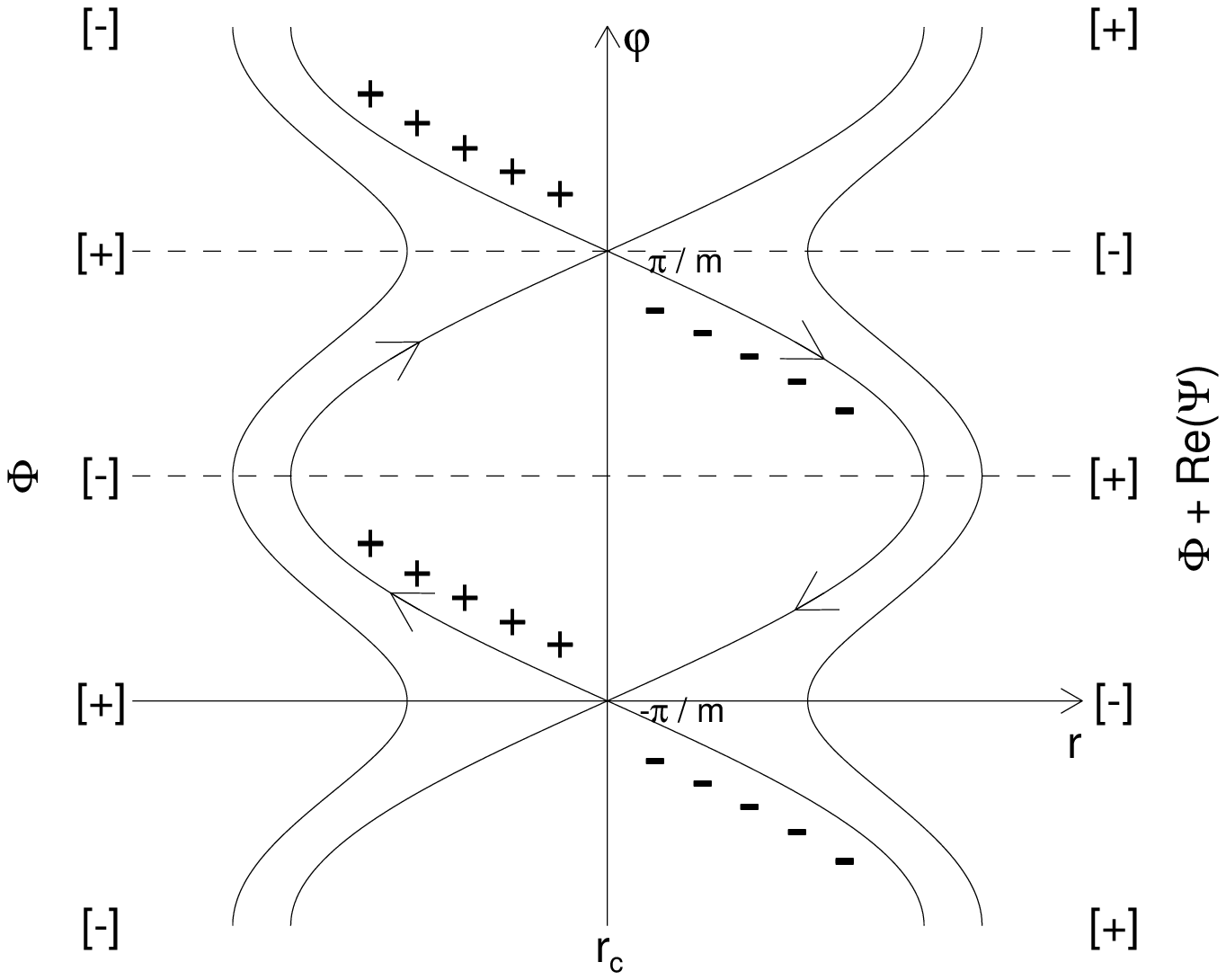} \figcaption{\label{fig:sketch}Sketch of
    the corotation region when $\Phi$ and $\Phi+\Re(\Psi)$ are in
    phase (left) and when $\Phi$ and $\Phi+\Re(\Psi)$ are in
    opposition (right).  The minima and maxima of $\Phi$ are indicated
    at the left, while the minima and maxima of $\Phi+\Re(\Psi)$ are
    indicated at the right. In the corotation region, material
    librates about the maxima of the effective potential
    $\Phi+\Re(\Psi)$. We assume a negative entropy gradient, hence
    material flowing outwards has a negative perturbed surface
    density, while material flowing inwards has a positive perturbed
    surface density, as indicated by the minus and plus signs.}
\end{figure}

The sign of this torque component can be understood as
follows. Fig.~\ref{fig:sketch} depicts the situation in two cases:
$\Phi$ and $\Phi+\Re(\Psi)$ have same sign (left), and $\Phi$ and
$\Phi+\Re(\Psi)$ have opposite signs (right).  In the left case, the
negative perturbed surface density on the outside of corotation is
located in the region where $\partial\varphi\Phi<0$, hence the 
perturbation yields a positive torque on the perturber. A similar conclusion
applies to the material flowing inwards which has positive
perturbation of surface density.  The torque on the perturber is therefore
positive, in agreement with the sign of $-{\cal S}\Phi[\Phi+\Re(\Psi)]$. 
An opposite conclusion holds for the case where $\Phi[\Phi+\Re(\Psi)] < 0$.

The order of magnitude and functional dependence of this torque
component can be justified as follows. As the sign has been justified
at the previous paragraph, we give here an estimate of the absolute
value. The perturbed surface density on the outside of corotation is
$\sim |{\cal S}\Sigma_0\delta/r_c|$, where
$\delta=[(\Phi+\Re(\Psi))/(-8AB)]^{1/2}$ is an order of magnitude of
the width of the libration islands. The specific torque in the region
of surface density perturbation is $\sim|m\Phi|$, while the area
covered by the perturbation of surface density scales with
$r_c^2\delta$. The torque arising from this region therefore scales
with $|mr_c\delta^2{\cal S}\Phi\Sigma_0|$, which is exactly the
scaling of $|\Gamma_0\,{\cal S}\Phi[\Phi+\Re(\Psi)]|$, within a
numerical factor in ${\cal O}(1)$.

The singular behavior of this torque component, which stems from
Eq.~(\ref{gcm2}), and which appears as a Dirac's delta function at
corotation, can be understood as follows: as the strength of the
perturbation decreases, the width of the libration islands tends to
zero, while the libration time tends to infinity (libration
disappears), hence we are left, in the linear regime, with a torque
contribution that comes strictly from the corotation radius and
therefore appears as singular.

It is worth noting that only half of the second term of
Eq.~(\ref{gcm}) comes from Eq.~(\ref{gcm2}). Eq.~(\ref{gcm1}), which
is obtained from the momentum flux jump, and which as such captures
effects occurring at a finite (albeit small) distance from corotation,
also displays a term similar to that of Eq.~(\ref{gcm2}). The
advection of entropy perturbations is not a silent process: it
triggers the emission of pressure waves \citep{ft00}. Our torque
expression indicates that half of the energy required to advect
entropy in the libration islands is evacuated through pressure waves.

\section {Application to the case of an embedded protoplanet}
\label{sec:planet}

In section~\ref{sec:isores}, we derived an expression for the
corotation torque between a radiatively inefficient disk and an
external rotating potential.  This expression is successfully
reproduced by local numerical simulations of an isolated corotation
resonance, in the linear regime.  We now contemplate the case of an
embedded protoplanet in a radiatively inefficient two-dimensional disk, 
for which all co-orbital corotation resonances are simultaneously active.

\subsection {Numerical features and setup}
\label{sec:plasetup}

Our numerical simulations are performed with the code {\sc Fargo}.  It
is a staggered mesh hydrocode that solves the Navier-Stokes,
continuity and energy equations on a polar grid.  It uses an upwind
transport scheme with a harmonic, second-order slope limiter
\citep{vl77}.  Its particularity is to use a change of rotating frame
on each ring of the polar grid, which increases the timestep
significantly \citep{fargo1,fargo2}, thereby lowering the
computational cost of a given calculation.  The energy equation that we 
implemented in {\sc Fargo} is:
\begin{equation}
  \frac{\partial e}{\partial t} + {\bf \nabla.}(e{\bf v}) =
  -p{\bf \nabla.v} + Q,
\label{eqnenergy}
\end{equation}
where $e$ is the thermal energy density, ${\bf v} = (u,r\Omega)^T$
denotes the flow velocity, $p$ is the vertically integrated pressure
and $Q$ is a heating source term that accounts for the disk viscosity
\citep[see e.g.][]{dangelo03b}.  The energy equation solver is
implemented as in \citet{zeus}.

In this work, the disk is taken inviscid so $Q=0$.  There is no
radiative transfer either, since the disk is assumed to be radiatively
inefficient.  Furthermore, $p$ and $e$ are connected by an ideal
equation of state $p = (\gamma - 1)e$, where the adiabatic index
$\gamma$ is set to $1.4$.  This equation of state can be expressed in
terms of the disk temperature $T$ and surface density $\Sigma$ as
$p=\Sigma T$.  The adiabatic sound speed reads $c_s = \sqrt{\gamma
  T}$, hence $c_s = \sqrt{\gamma}\,c_{s,\rm{iso}}$, where $c_{s,
  \rm{iso}}$ refers to the isothermal sound speed. We comment that the 
Lindblad torque, which scales as $c_s^{-2}$ \citep{w97}, is therefore 
weakened by a factor of $\gamma$ in an adiabatic disk. The same is true of 
the corotation torque, when there is no entropy gradient. We checked both 
effects with appropriate calculations, not reproduced here. 
This plays in favor of a total torque  reversal in adiabatic disks 
with a negative entropy gradient.

The disk is initially slightly sub-Keplerian (the pressure gradient is
accounted for in the centrifugal balance), axisymmetric, with
power-law profiles for the surface density and temperature given by
Eqs.~(\ref{eqn:siglaw}) and~(\ref{eqn:tlaw}).

For a comparative purpose, calculations involving a locally isothermal
equation of state are performed. In isothermal calculations, no energy
equation is solved: $p$ and $\Sigma$ are simply connected by $p =
\Sigma c^2_{s, \rm{iso}}$. These isothermal calculations have same 
initial surface density and temperature profiles as the adiabatic runs.

The disk is perturbed by the smoothed potential of a protoplanet.
We adopt a Plummer potential, with a softening length 
$\varepsilon = 0.6 H(r_p)$ (unless otherwise stated), $r_p$ being 
the planet orbital radius. This fiducial
value is quite substantial for our purposes, but investigating the
disk response at much smaller softening lengths, where the adiabatic
effects on the corotation torque are increasingly important, requires
a very large resolution. A high resolution systematic study at small
softening length will be presented in a forthcoming work.

The protoplanet is held on a fixed circular orbit, at $r=r_p$. The disk
parameters are summed up in Table~\ref{param}, where they are
expressed in the following unit system: $r_p$ is the length unit, 
the mass of the central object
$M_*$ is the mass unit and $(GM_* / {r_p}^3 )^{-1/2}$ is
the time unit, $G$ being the gravitational constant ($G = 1$ in our
unit system).  We denote by $T_{\rm orb}$ the planet orbital period, 
$M_p$ the planet mass and $q=M_p/M_*$ the planet to primary mass ratio.
\begin{deluxetable}{lcc}
  \tabletypesize{\scriptsize} \tablecaption{Reference parameters.  The
    disk is inviscid
    \label{param}} \tablewidth{0pt} \tablehead{ \colhead{Parameter} &
    \colhead{Notation} & \colhead{Reference value} } \startdata
  Aspect ratio at $r=r_p\dotfill$ & $h(r_p)$ & $0.05$\\
  Surface density at $r=r_p\dotfill$ & $\Sigma_p$ & $2\times10^{-3}$\\
  Softening length$\dotfill$ & $\varepsilon$ & $0.03$\\
  Adiabatic index$\dotfill$ & $\gamma$ & $1.4$\\
  Mesh inner radius$\dotfill$ & $r_{\rm min}$ & $0.4$\\
  Mesh outer radius$\dotfill$ & $r_{\rm max}$ & $1.8$\\
  Radial zones number$\dotfill$ & $N_r$ & $512$\\
  Azimuthal zones number$\dotfill$ & $N_s$ & $2048$\\
\enddata
\smallskip
\end{deluxetable}
\subsection {Results}

\subsubsection {An illustrative example}
\label{sec:example}
\begin{figure}
  \plottwo{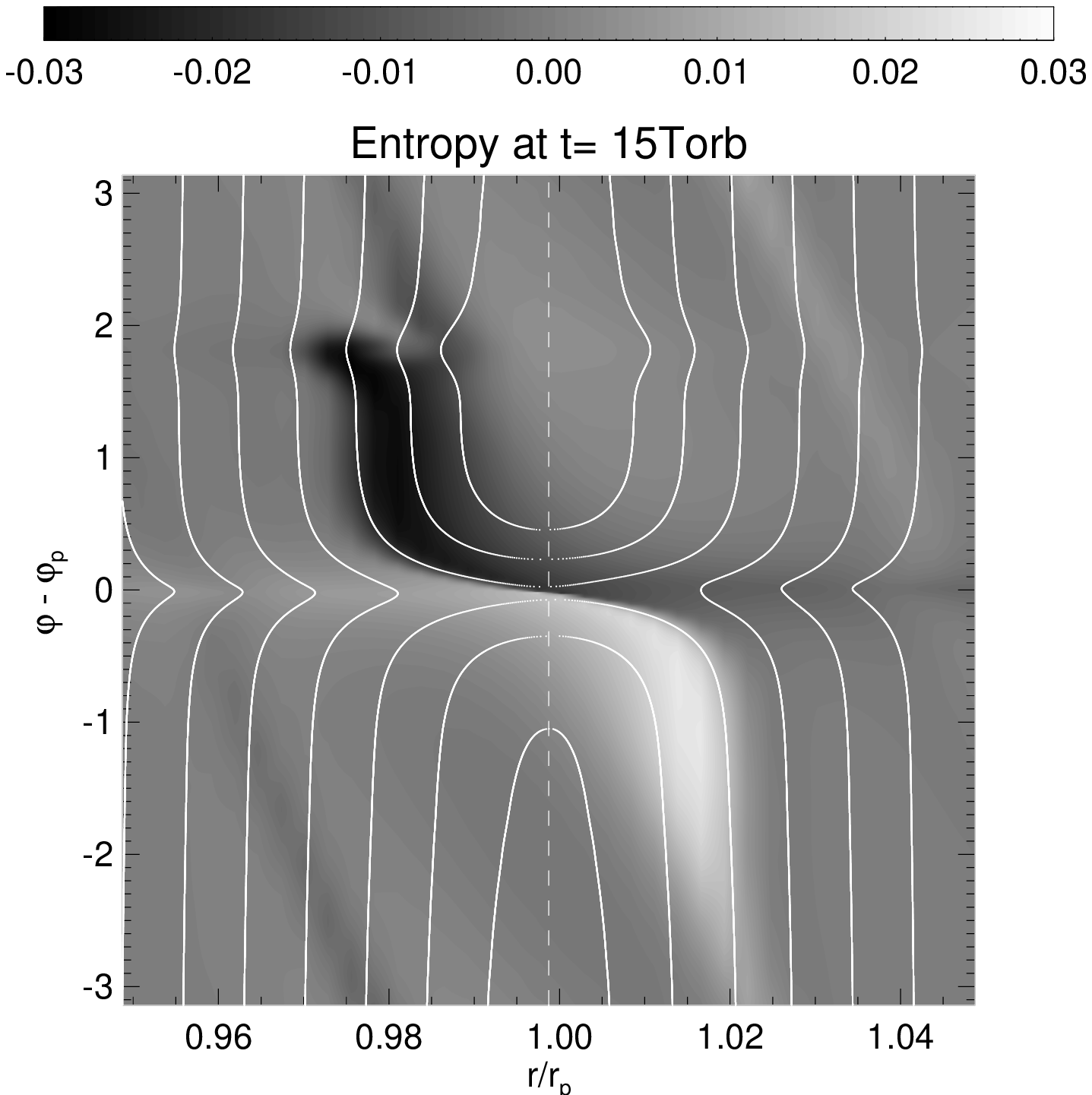}{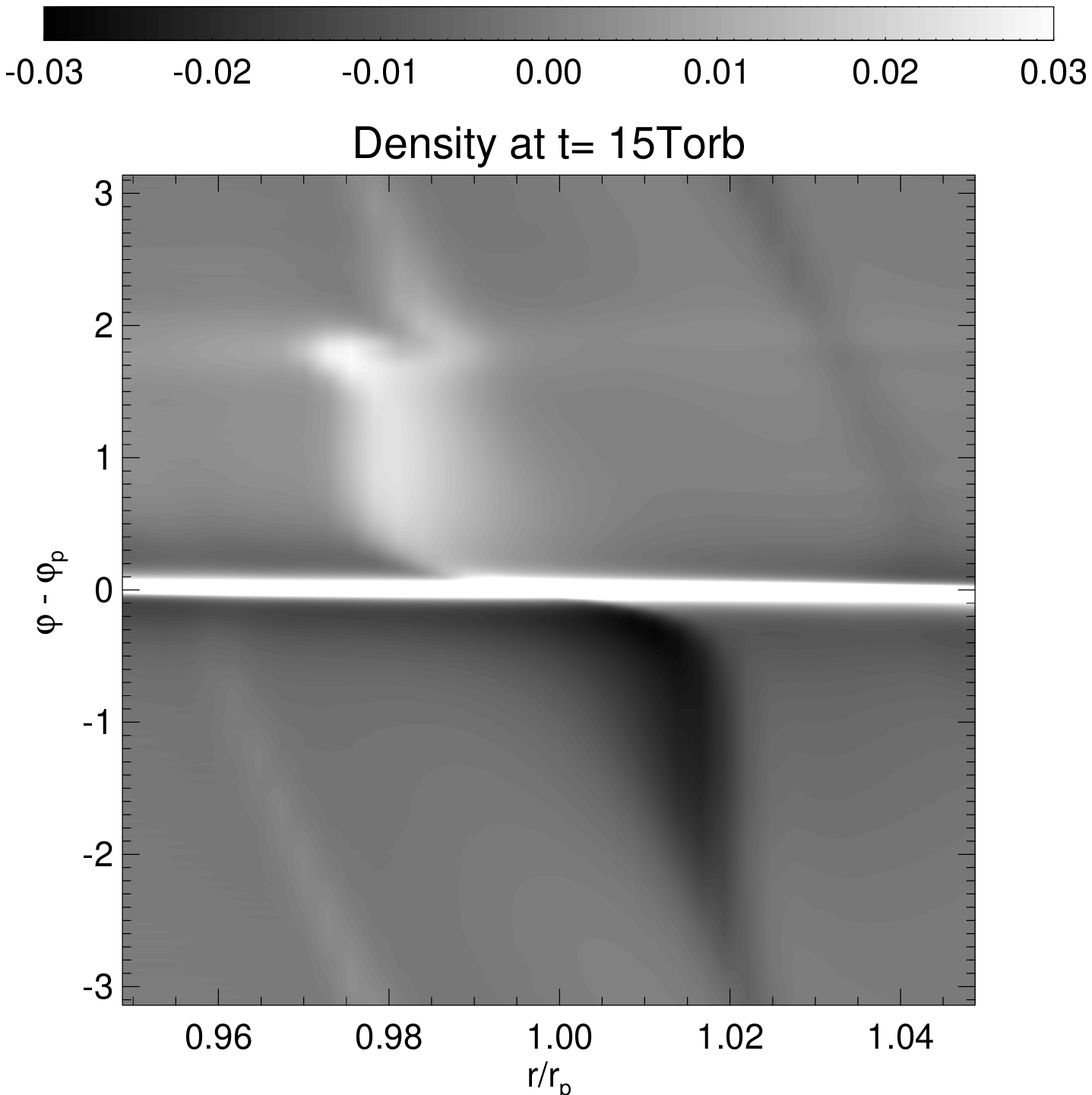} \plottwo{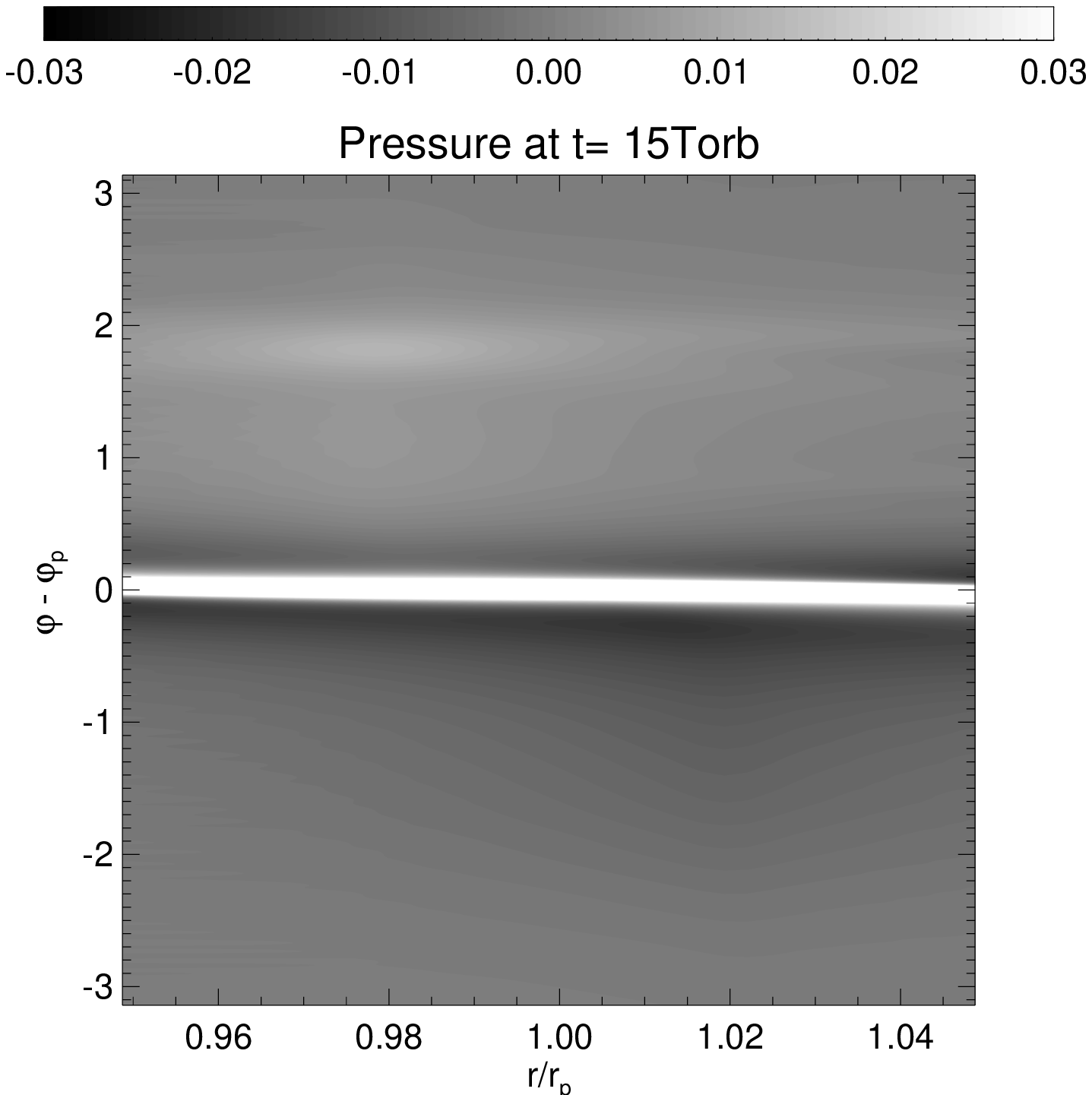}{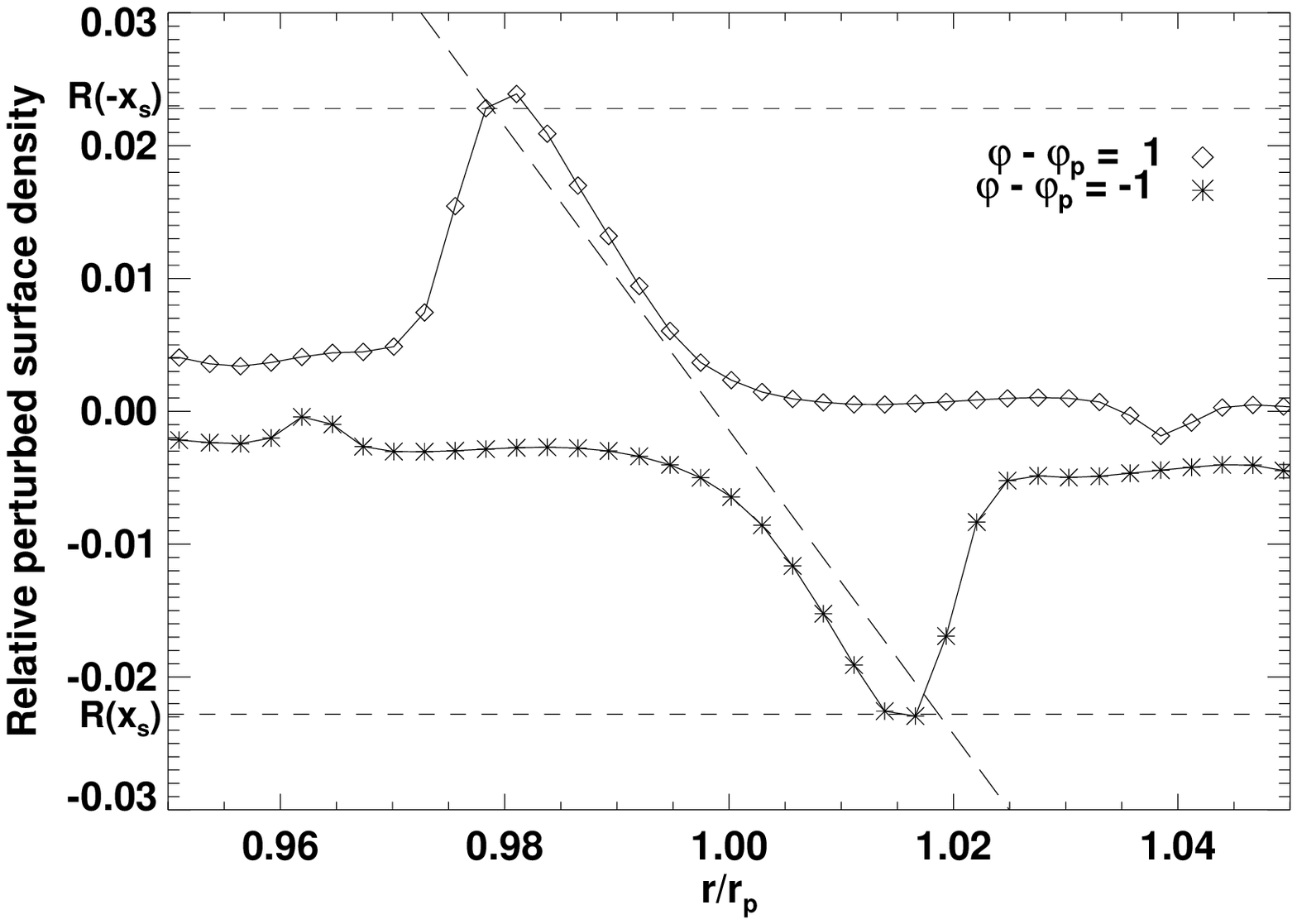}
  \figcaption{\label{medley}Top row and bottom left: relative
    perturbations of the gas entropy, surface density and pressure, at
    $t = 15\,T_{\rm orb} \approx \tau_{\rm lib} / 4$. The protoplanet
    is located in $r=r_p$, $\varphi=\varphi_p$. In the top left panel,
    streamlines are overplotted and the vertical dashed line stands
    for the corotation radius $r_c$.  In the top right and bottom left
    panels, the color scale is adjusted to highlight the advection of
    the entropy perturbation (see text). The nearly horizontal
    overdensity structure at $\varphi = \varphi_p$ is the
    protoplanet's wake. Bottom right: slices of the relative perturbed
    density field at the same time, at $\varphi-\varphi_p = 1$
    (diamonds) and $\varphi-\varphi_p = -1$ (stars). The two
    horizontal dashed lines refer to the values of $R(-x_s)$ and
    $R(x_s)$, while the long-dashed curve displays the quantity
    $2(r-r_c){\cal S}/r_c$ (see text and Eq. (\ref{rplus2})).}
\end{figure} 

We show the results of an illustrative calculation with a $q=2.2\times
10^{-5}$ planet to primary mass ratio (corresponding to
$M_p=7.3\;M_\oplus$ if the central object has a solar mass).
The horseshoe libration time is
\begin{equation}
  \label{eq:hslib}
  \tau_{\rm lib} = \frac{8\pi r_p}{3\Omega_p x_s},
\end{equation}
where $\Omega_p$ is the protoplanet angular velocity and $x_s$ denotes
the half-width of the horseshoe region.  \citet{mak06} have given an
estimate of $x_s$ in the isothermal case, that reads $x_s \approx 1.16
r_p \sqrt{q/h(r_p)}$.  A streamline analysis was performed and
confirmed that this estimate holds for an adiabatic disk, if one
substitutes $h(r_p)$ with $\sqrt{\gamma} h(r_p)$.  We find therefore
$\tau_{\rm lib} \approx 60\,T_{\rm orb}$. Numerical diffusion eventually 
alters the conservation of entropy. Nevertheless, the horseshoe region 
spans $20$ zones radially, which is sufficient to follow the horseshoe 
dynamics over several libration times. Since we are concerned here with 
a fraction of the libration time, the entropy is conserved with a good 
level of accuracy over the duration of our runs, and it can be regarded 
as a Lagrangian tracer of the flow.

Two calculations were performed: an adiabatic and an isothermal one.
Both lasted thirty orbital periods, hence half the horseshoe libration
time.  This calculation has $\sigma=0.5$ and $f=0$, as in PM06.  This
gives ${\cal S}\approx -0.57$.

Fig.~\ref{medley} displays the gas entropy, surface density and
pressure obtained in the adiabatic calculation, after $15\,T_{\rm orb}$. 
Each field represents the relative perturbation of the
corresponding quantity with respect to the unperturbed state. For
instance, the top right panel shows $[\Sigma (r,\varphi) -
\Sigma_0(r)] / \Sigma_0 (r)$.  While the azimuthal range spans the
whole $[0,2\pi]$ interval, the radial range depicted is restricted to
a band of width $2.5 x_s$ around the corotation radius $r_c$. 
We overplot streamlines to the entropy panel to give an idea of the
extent of the horseshoe region. The vertical dashed line represents
the corotation radius.  Whereas the pressure panel does not display
any significant perturbation, the entropy and density panels show the
propagation of a perturbation inside the horseshoe region, which
slides along the separatrices. This is reminiscent of the behavior
commented in the case of an isolated resonance at
section~\ref{sec:dyncor}.

The interpretation of this dynamics is as follows: the entropy of the
fluid elements is conserved as they perform a horseshoe U-turn in the
co-orbital region. When there is initially an entropy gradient at
corotation, the co-orbital dynamics yields an entropy perturbation that
has a sign opposite of that of the entropy gradient on the outwards
U-turns, and the sign of the entropy gradient on the inwards U-turns.
Since the pressure field is only weakly perturbed, the entropy
perturbation is related to a density perturbation of opposite sign
and, in relative value, of same order of magnitude. Therefore, if
there is a negative entropy gradient at corotation (${\cal S}<0$, as
in the example shown here), the co-orbital dynamics yields a negative
density perturbation at $\varphi < \varphi_p$ and a positive density
perturbation at $\varphi > \varphi_p$, with straightforward consequences 
for the corotation torque.
Using an expression inherited from the terminology of Riemann
solvers, we call this perturbation a contact discontinuity.  A contact
discontinuity is characterized by a discontinuity in the density and
temperature fields, while the pressure and velocity fields are
continuous. A contact discontinuity is simply advected by the
flow. Here it follows the horseshoe dynamics, and it remains confined
to the horseshoe region.

We give hereafter a simple estimate of the relative perturbation of
the disk surface density due to the advection of entropy.  We consider
a fluid element that performs a horseshoe U-turn from the inner part
of the horseshoe region (where we assume that there is no entropy
perturbation, which is true as long as $t < \tau_{\rm lib}/2$) to the
outer part.  All physical quantities at the inner (outer) leg of the
horseshoe streamline are denoted by a minus (plus) subscript.  A
first-order expansion yields, assuming no pressure perturbation:
\begin{equation}
  p_{\pm} = p_0(r_c) (1\mp\lambda x/r_c),
\end{equation}
where $0<x<x_s$ is the distance of the streamline to corotation, and:
\begin{equation}
  \Sigma_{-} = \Sigma_0(r_c) (1+\sigma x/r_c).
\end{equation}  
On the outer horseshoe leg, the disk surface density is perturbed
according to the entropy perturbation and reads:
\begin{equation}
  \Sigma_{+} = \Sigma_0(r_c) (1+R-\sigma x/r_c),
\end{equation}
where $R$ is the relative perturbation of surface density at $r=r_c +
x$ (we assume a symmetric horseshoe U-turn), due to the entropy
advection.  Entropy conservation along the fluid element path ($S_{-}
= S_{+}$) leads to:
\begin{equation}
  R = 2\frac{x}{r_c}\left(\sigma - \frac{\lambda}{\gamma}\right) =
  2\frac{x}{r_c}{\cal S}.
\label{rplus}
\end{equation}
The horseshoe U-turn that we have considered lags the planet ($\varphi
<\varphi_p$).  A similar conclusion holds for a horseshoe U-turn
that switches from the outer leg to the inner one (at
$\varphi>\varphi_p$), hence we finally have: 
\begin{equation}
  R(x) = 2x{\cal S}/r_c, \,\forall x \in [-x_s, +x_s].
  \label{rplus2}
\end{equation}  

The bottom right panel of Fig.~\ref{medley} displays the slices of
the perturbed density field at $t=15\,T_{\rm orb}$, for
$\varphi-\varphi_p = 1$ (diamonds) and $\varphi-\varphi_p =
-1$ (stars).  The two horizontal dashed lines display the values
of $R(-x_s)$ and $R(x_s)$, where $x_s$ is estimated through a
streamline analysis.  Similarly, the long-dashed curve shows
$R(x)=2x{\cal S}/r_c$, which is in correct agreement with the
calculation results.  The surface density structure in the horseshoe
region is therefore dictated by the sign of ${\cal S}$.  In
particular, we do not expect any contact discontinuity in the
homentropic case (${\cal S}=0$). We have checked this prediction with
a numerical simulation (not presented here).

\subsubsection {Excess of corotation torque and entropy gradient}
\label{sec:sgrad}

\begin{figure}
  \plotone{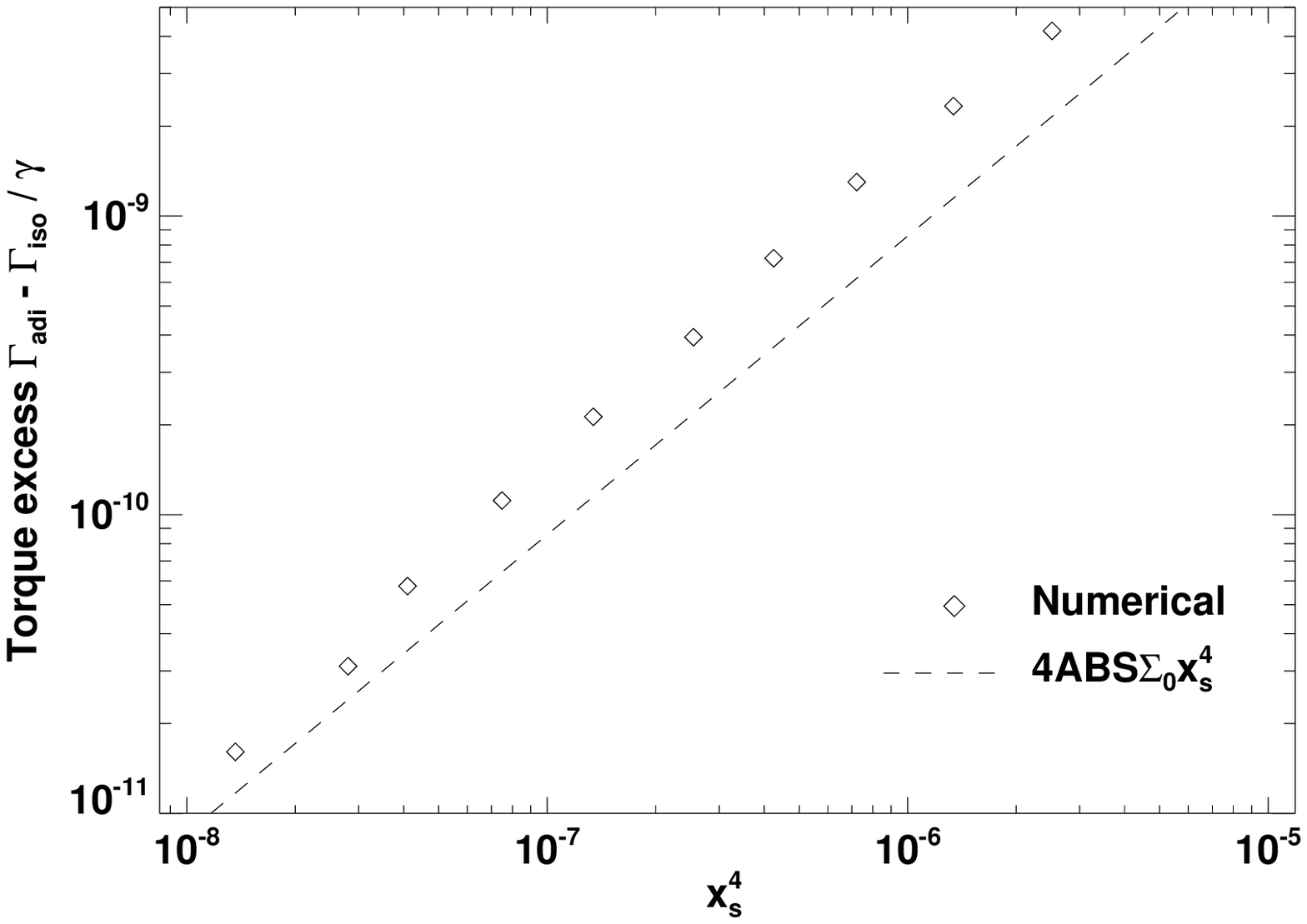} \figcaption{\label{gcrxs4}Torque excess (see text)
    as a function of the half-width of the horseshoe region.}
\end{figure}

An order of magnitude of the excess of corotation torque arising from the
perturbation of the surface density field can be given by estimating
how the standard horseshoe drag expression \citep{w91,m01} is modified
by the perturbation of surface density $R(x)\Sigma_0(r_c)$.  We
consider the outwards horseshoe U-turns that occur at
$\varphi<\varphi_p$. Assuming, in this order of magnitude estimate,
that the rotation profile of the disk is unperturbed, we evaluate the
variation of angular momentum flux of the horseshoe disk material
after the U-turn attributable to the change of the disk's surface
density:
\begin{equation}
  \label{eqn:hsdm}
  \Delta\Gamma_{\rm HS^-}=\int_0^{x_s}(-2Ax)\Sigma_0R(x)(j_c+2Br_cx)dx,
\end{equation}
where $j_c$ is the specific angular momentum of the material at corotation.
The first factor of the integrand of Eq.~(\ref{eqn:hsdm}) represents
the material velocity in the corotating frame, due to the shear. The
last factor is the material specific angular momentum obtained from a
first order expansion at corotation.
Similarly, we obtain the change of angular momentum flux due to the
perturbation of surface density on inwards horseshoe U-turns:
\begin{equation}
  \label{eqn:hsdp}
  \Delta\Gamma_{\rm HS^+}=\int_0^{x_s}(-2Ax)\Sigma_0R(-x)(j_c-2Br_cx)dx.
\end{equation}
Adding Eqs.~(\ref{eqn:hsdm}) and~(\ref{eqn:hsdp}), we are left with:
\begin{equation}
  \Delta\Gamma_{\rm HS}=2\int_0^{x_s}(-2Ax)\cdot \Sigma_0R(x)\cdot 2Br_cxdx
  =-4AB\Sigma_0{\cal S}x_s^4.
\label{eqn:hsd}
\end{equation}
Fig.~\ref{gcrxs4} shows the excess of corotation torque between an
adiabatic and isothermal calculation with same parameters, as a function 
of the half-width of the horseshoe region.
This excess is obtained by subtracting the total torque of an adiabatic and an
isothermal calculation (the isothermal torque being rescaled by a factor 
$\gamma^{-1}$, since $c_s=\sqrt{\gamma}\,c_{s,\rm iso}$).
We call this difference the torque excess for further reference.
Each data point corresponds to a 
calculation with a given planet mass, for which we determine $x_s$
through a streamline analysis. We find that the torque excess
approximately scales as $x_s^4$, and that it is within a factor~$2$ of
our order of magnitude estimate, given by $-\Delta\Gamma_{\rm HS}$.

The torque expression of Eq.~(\ref{gcm}) as well as the horseshoe drag
expression of Eq.~(\ref{eqn:hsd}) suggest that the torque excess scales 
with ${\cal S}$, hence with the entropy gradient. 
In order to check that, we have undertaken a number of calculations 
with different values of ${\cal S}$. 
These calculations have $q=2.2\times 10^{-5}$,
and the disk parameters are those of Table~\ref{param}. Each entropy
gradient is realized with different combinations of the indexes of
the pressure and surface density power laws. Adiabatic effects on 
the torque are assessed in two different ways:
\begin{enumerate}
\item By calculating the torque excess, as in Fig.~\ref{gcrxs4}.
\item By evaluating the following integral:
\begin{equation}
  \Gamma_{\rm cd}=\int_{\rm disk}\left(\Sigma
    -\frac{p}{c_s^2}\right)\frac{\partial\Phi}{\partial\varphi}rdrd\varphi,
\label{eqn:cd}
\end{equation}
which provides an estimate of the torque due to the contact
discontinuity (this contribution arises from perturbations of $\Sigma$
which do not have a pressure counterpart). In the linear regime,
Eq.~(\ref{eqn:cd}) amounts to a summation over $m$ of the last term of
Eq.~(\ref{s1approx}). We shall check this statement in the next
section.
\end{enumerate}

\begin{figure}
  \plottwo{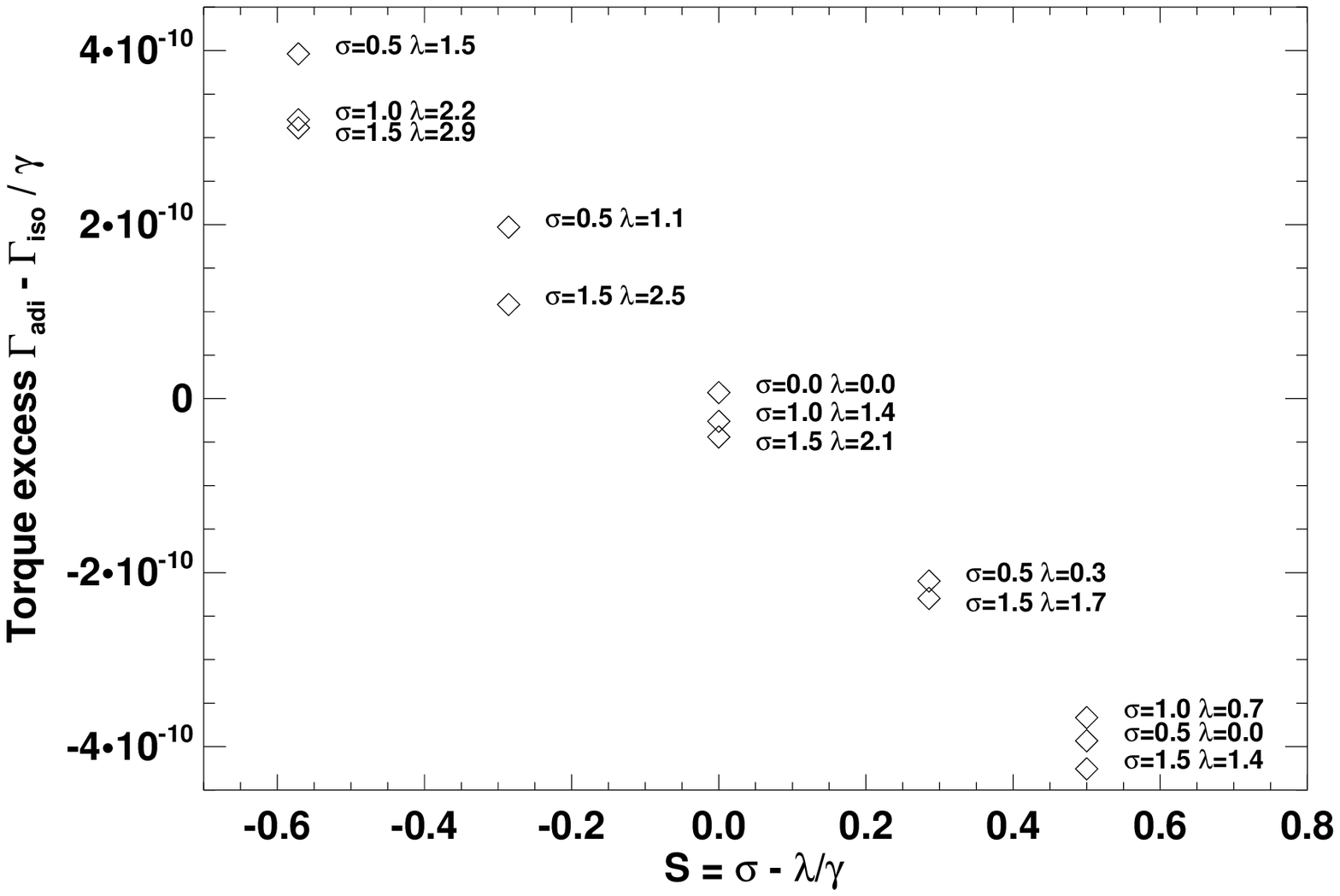}{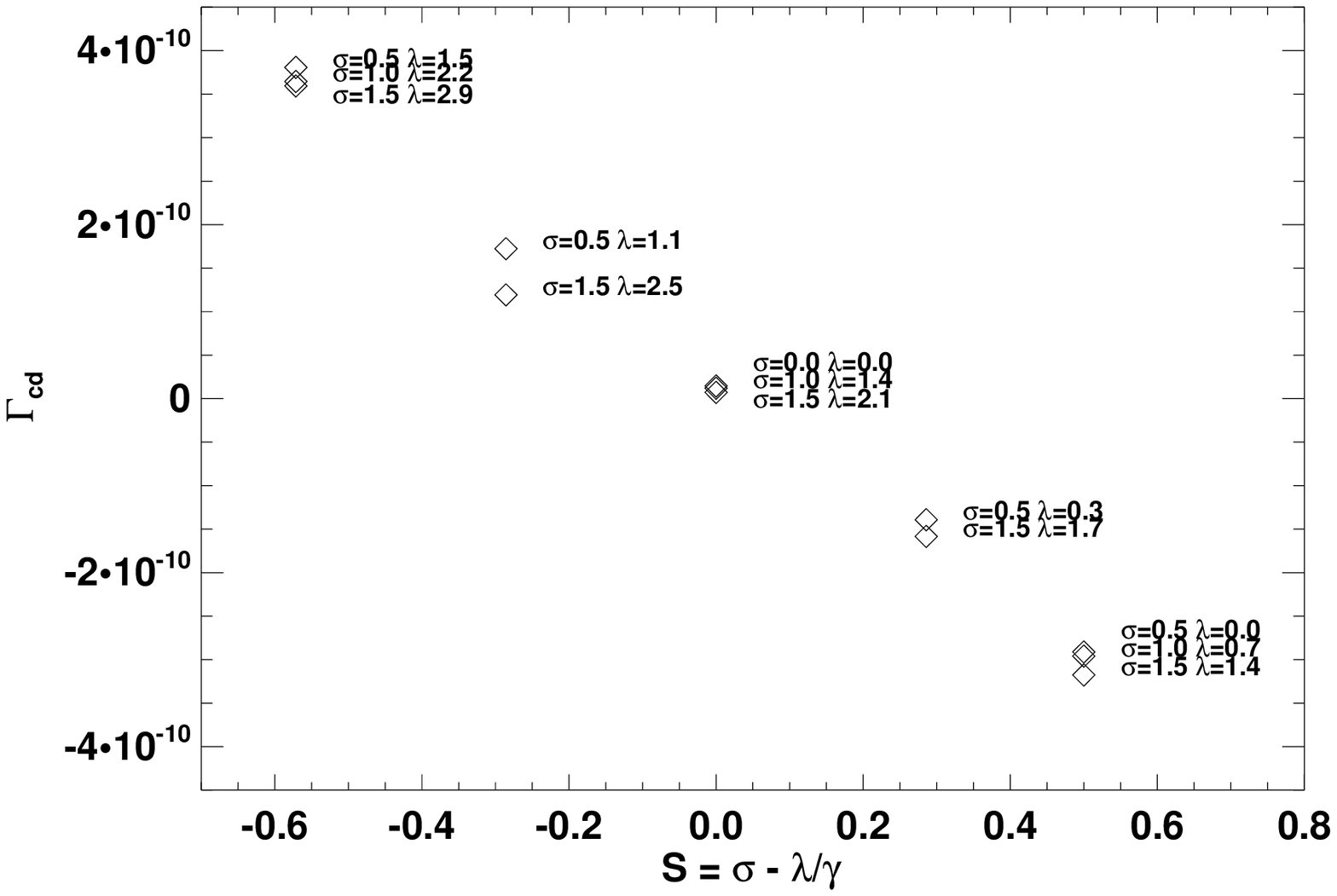} \figcaption{\label{tqexcess}Torque excess
    (left) and contact discontinuity contribution to the torque
    (right) as a function of ${\cal S}$.  Although the calculations
    display some scatter for a given value of ${\cal S}$, the
    different points can be considered as aligned within a good level
    of approximation. The slope of the dependence is negative.}
\end{figure}

These two estimates of adiabatic effects on the torque value are shown
respectively in Figs.~\ref{tqexcess}a and~\ref{tqexcess}b. Remarkably,
they coincide within $\sim 25$~\%. We will comment further this
coincidence in the next section.

The main conclusion that can be drawn from the results of
Fig.~\ref{tqexcess} is that the torque excess (or the contact
discontinuity contribution) essentially depends on the entropy
gradient, as expected. The excess is positive for a negative entropy
gradient, hence we may expect the total torque exerted on a planet
embedded in a radiatively inefficient disk to be a positive quantity
if the radial entropy gradient is sufficiently negative.

\subsubsection {Connection to the analytical expression}
\label{sec:connect}

\begin{figure}
  \plotone{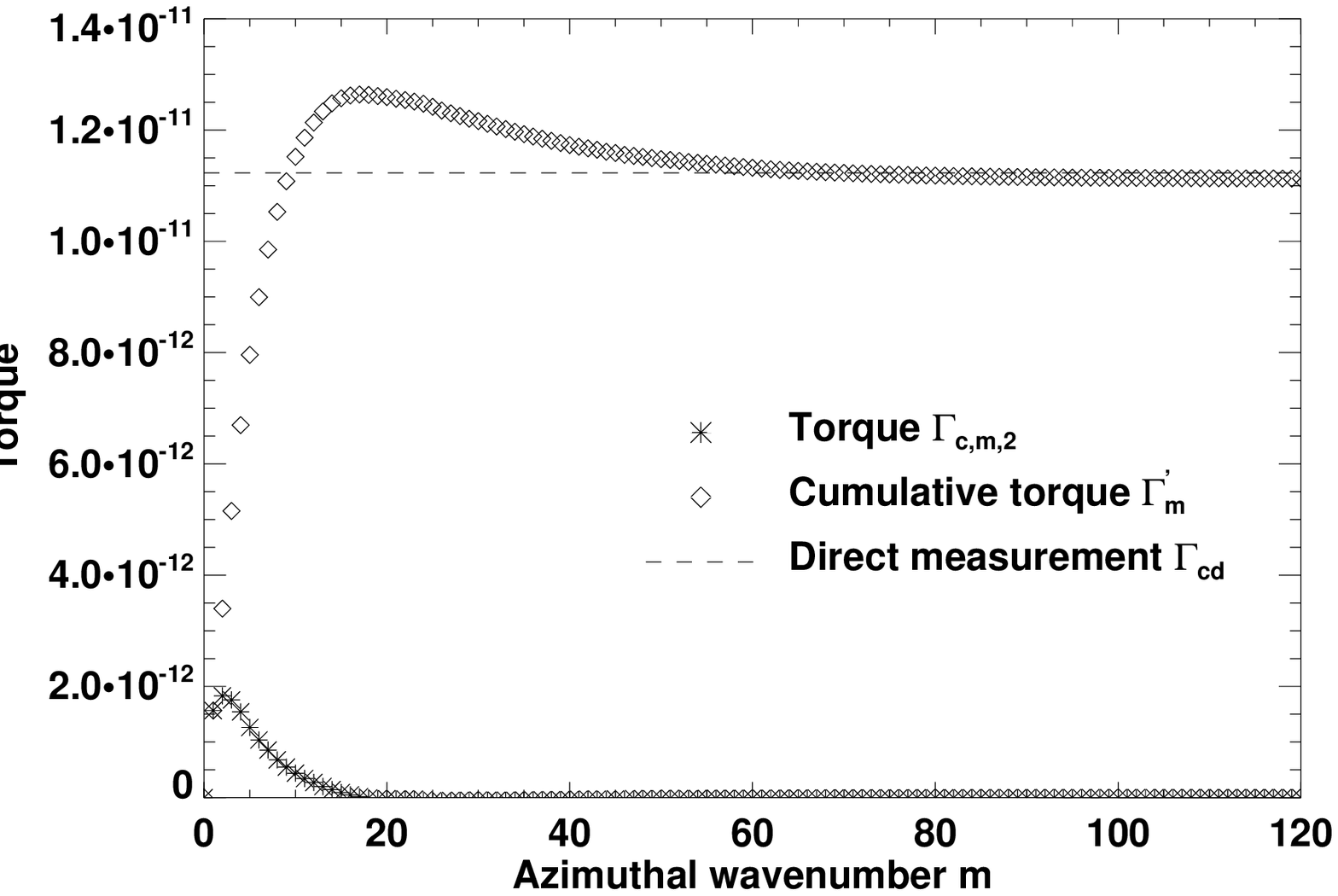} \figcaption{\label{fig:accord}Partial sums of the
    torque series given by Eq.~(\ref{eqn:partial}) (diamonds) and
    direct estimate of the contact discontinuity contribution, given
    by Eq.~(\ref{eqn:cd}) (dashed line). The asymptotic value of the
    partial sum almost coincides with the direct estimate (i.e. the
    diamonds almost lie on the dashed line at large $m$), hence with a
    very good accuracy we have $\Gamma_\infty=\Gamma_{\rm cd}$ (see
    text).}
\end{figure}

We have given at Eq.~(\ref{gcm2}) an estimate of the singular torque
contribution from the contact discontinuity at an isolated resonance,
while we have estimated the total contribution in the planetary case
of the contact discontinuity using Eq.~(\ref{eqn:cd}) at
section~\ref{sec:sgrad}. We check in the present section that this total
contribution corresponds to the sum over $m$ of the torque expression
of Eq.~(\ref{gcm2}). For this purpose, we have adopted a planet to primary mass
ratio $q=5\times 10^{-6}$, as the one adopted in the previous sections
($q=2.2\times 10^{-5}$) led to poor agreement, presumably because of
the onset of non-linear effects. For each azimuthal wavenumber $m$, we
measure $\Re(\Psi_m)$ from the calculation output (at $t = 5\,T_{\rm orb}$),
and we evaluate the sum over $m$ of the torque $\Gamma_{c,m,2}$:
\begin{equation}
  \Gamma_{\infty}=\lim_{k\rightarrow +\infty}\Gamma'_k,
\label{eqn:sommation}
\end{equation}
where:
\begin{equation}
  \label{eqn:partial}
  \Gamma'_k=-\frac{4\pi^2}{3}\left[\frac{{\cal
        S}\Sigma_0}{\Omega^2}\right]_{r_c}
  \,\sum_{m=1}^{m\le k}m\Phi_m[\Phi_m+\Re(\Psi_m)]
\end{equation}
is the partial sum of $\Gamma_{c,m,2}$. We compare the torque
contribution given by Eq.~(\ref{eqn:cd}) to $\Gamma_{\infty}$. The
results are presented in Fig.~\ref{fig:accord}.  The agreement between
the direct torque measurement and the linear estimate is excellent.

\begin{figure}
  \plottwo{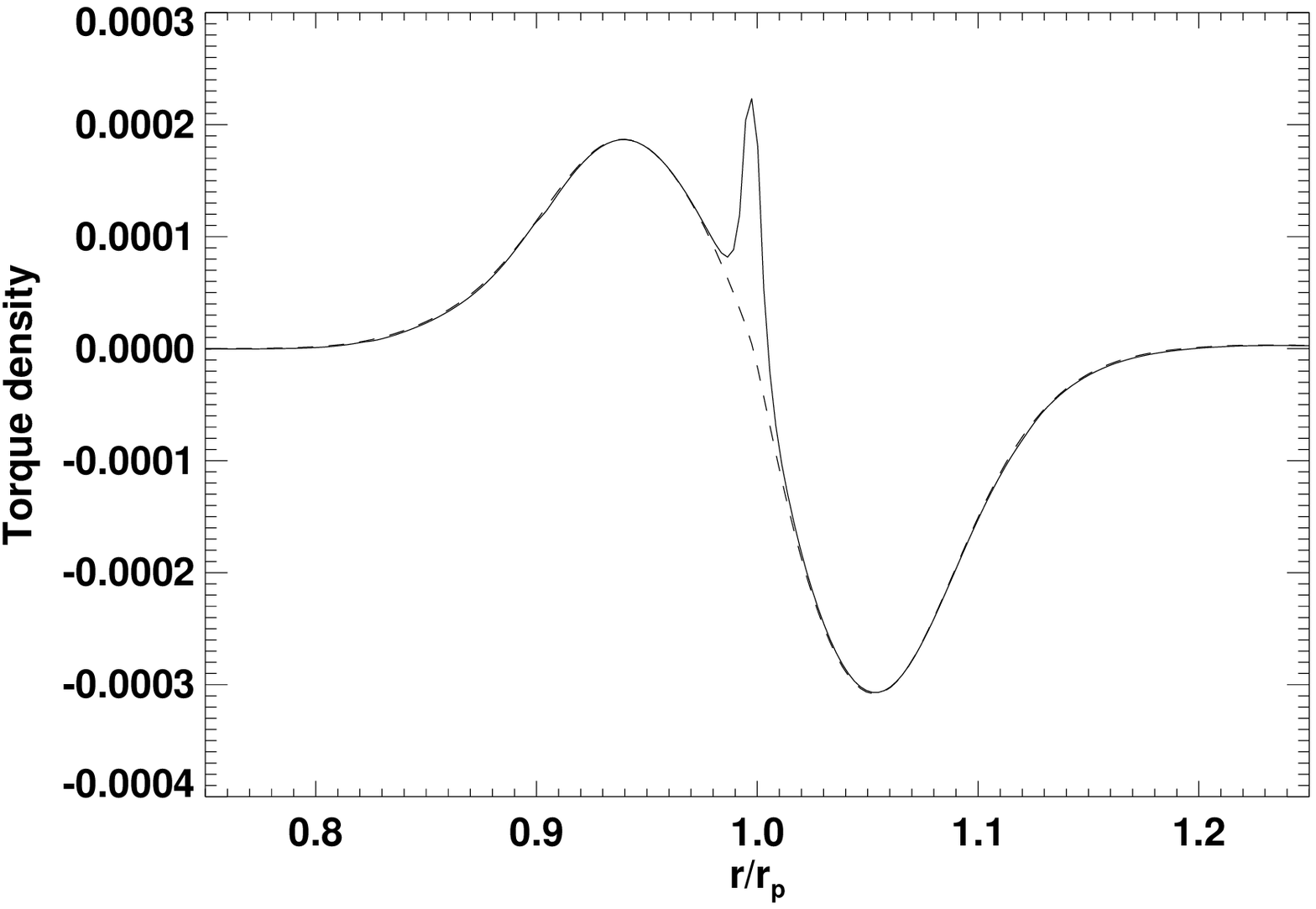}{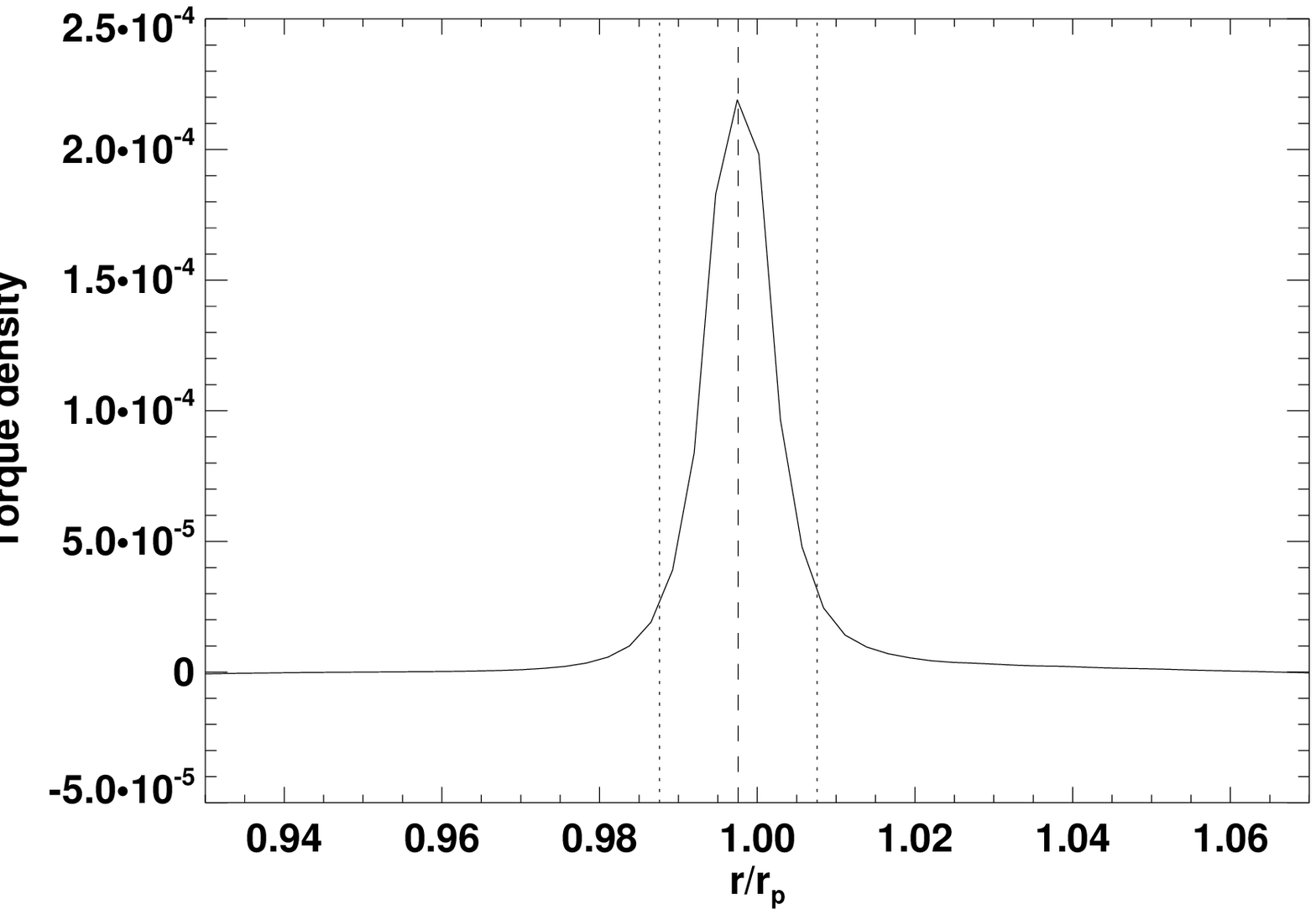} \figcaption{\label{tqdens}Left: total
    torque density (solid curve) and torque density of $p/c_s^2$
    (dashed curve). Right: torque density of $\Sigma-p/c_s^2$. The
    vertical dashed line shows the corotation radius, while the two vertical
    dotted lines show the extent of the horseshoe region.}
\end{figure}

This confirms what we anticipated in section~\ref{sec:cortq}, and what
is shown in appendix~\ref{apA}, that the contribution of the last term
of Eq.~(\ref{ims1approx}) to the torque is negligible in the planetary
context. Also of interest is the torque density associated
respectively to $p/c_s^2$ and $\Sigma-p/c_s^2$. The sum of these two
torque densities is the total torque density. They are represented at
Fig.~\ref{tqdens}. The total torque density displays a smooth profile
and a narrow peak at corotation. This is reminiscent of the torque
density found by PM06 (their Fig.~$2$) or by \citet{mota03} (their
Fig.~$3$). The decomposition above splits this total torque density in a
smooth component arising from $p/c_s^2$, which reminds the torque
density in an isothermal disk, and a sharp, localized torque density
arising from $\Sigma-p/c_s^2$.  This corresponds to the torque density
of the contact discontinuity contribution given by
Eq.~(\ref{eqn:cd}). Fig.~\ref{tqdens} shows that this contribution
(which is singular at corotation in the linear case for an isolated
resonance) is here bounded by the extent of the horseshoe region.

We comment the surprising agreement found at the previous section
between the torque excess and the contribution of the contact
discontinuity.  The linear analysis suggests that the former should be
the sum of $\Gamma_{0,m}\,[2{\cal S}|\Phi_m+\Psi_m|^2 
- 2{\cal S}\Phi_m(\Phi_m+\Re(\Psi_m))]$, which,
in the limit where $|\Phi_m+\Re(\Psi_m)| \ll |\Re(\Psi_m)|$ and
$|\Phi_m+\Re(\Psi_m)| \ll |\Phi_m|$, should reduce to
$-2\Gamma_{0,m}\,{\cal S}\Phi_m(\Phi_m+\Re(\Psi_m))$, that is twice 
the contribution of the contact discontinuity (see section~\ref{sec:gen}). 
Nevertheless, for the calculations presented
here, we can check that $2\sum_mm|\Phi_m+\Psi_m|^2$ is almost exactly
compensated by $\sum_mm\Phi_m(\Phi_m+\Re(\Psi_m))$. Namely the ratio
of the former to the latter quantity is found to be $1.07$, which
explains why the full excess expression essentially amounts to the
contact discontinuity contribution. Presumably this coincidence is
fortuitous and linked to the relatively large softening length that we
use.  At smaller softening length, the term in
$\Phi_m(\Phi_m+\Re(\Psi_m))$ should largely dominate, yielding a ratio
of $2$ between the torque excess and the contribution of the contact
discontinuity. We note that PM06 also quote that the torque estimate
given by their equation~(1) accounts for the total torque within
$25$~\% (this equation can also be seen as an estimate of the contact
discontinuity contribution). This seems to suggest that the softening
length of $0.6H(r_p)$ that we adopted is a correct choice to reproduce the
magnitude of the corotational effects in adiabatic three-dimensional
disks.

\section{Discussion}
\label{sec:discu}

\subsection{Softening length}
\label{sec:soften}

In an isothermal disk, the corotation torque scales with
$|\Phi+\eta|^2$ \citep{tanaka}. Even if $\Phi$ diverges at corotation,
$\Phi+\eta$ remains finite, which makes the isothermal corotation
torque a quantity relatively insensitive to the softening length. The
situation is quite different for the effects linked to the entropy
advection that we present here: they involve the product
$\Phi(\Phi+\Psi)$, which diverges when $\Phi$ does. Adiabatic effects
on the corotation torque should acquire a very large magnitude at
small softening length. In particular, if the softening length is
smaller than the distance from orbit to corotation, the magnitude of
these effects should strongly depend on this distance, which scales
with the pressure gradient. If one regards the softening length as a
proxy for the altitude in a three-dimensional disk, the extent of the
disk vertical scale height concerned by these very small softening
length issues should be small, however, since the distance from orbit
to corotation is a fraction of $r_p h^2$. Nevertheless, it is of
interest to investigate the behavior of the corotation torque in an
adiabatic flow at very small softening length to assess the importance
of such effects. Owing to the very large resolution required to
investigate this problem, we defer this investigation to a forthcoming
work.

\subsection{Saturation}
\label{sec:satur}

The origin of the effects presented here is the advection of entropy
in the corotation region, that triggers an entropy perturbation (and
therefore a density perturbation) whenever there is an entropy
gradient in the equilibrium profile. Libration occurs on different
timescales for the different streamlines of the corotation region,
which tends to stir the entropy and to flatten out the entropy profile
across the corotation region (be it the horseshoe region in the
planetary case or a libration island in the isolated resonance case).
This is quite similar to the behavior of the corotation torque in an
isothermal disk, which tends to saturate because the vortensity
profile is flattened out by libration. In this case, it is the viscous
diffusion which can prevent the flattening out of the profile if it
acts sufficiently rapidly to establish the large scale gradients
before a libration time.  This has been studied for an isolated
resonance by \citet{gs03} and~\cite{ol03}, and by \citet{bk01}
and~\citet{m01} for a planetary co-orbital region. In both cases, the
degree of saturation of the corotation torque in steady state depends
on the ratio of the libration time and of the viscous time across the
libration region. The dissipative processes required to prevent the
torque saturation in the situation presented here should be able to
impose the large scale entropy gradient over the corotation region in
less than a libration time. Radiative processes (cooling and heating)
should therefore occur on a timescale longer than a horseshoe U-turn
(otherwise the flow can rather be considered as locally isothermal),
but they should act on a timescale shorter than the libration time.
We provide an estimate of the horseshoe U-turn time and of the
libration time for a small mass object embedded in a gaseous disk. The
horseshoe half-width $x_s$ is $\sim r_p\sqrt{q/h(r_p)}$.  
Neglecting pressure effects
and writing a simplified Jacobi constant for a test particle near a
horseshoe U-turn as: $J=-GM_p/(2Br_p|\varphi-\varphi_p|)+A(r-r_p)^2$, we can
estimate the distance of closest approach between the planet and a
test particle flowing along a horseshoe separatrix as
$r_p|\Delta\varphi|_s=\Omega_p^2\,H(r_p)/2|AB|={\cal O}(H(r_p))$. 
The time required to
perform a horseshoe U-turn can be deduced using the radial drift
velocity of the test particle when it crosses the orbit, at its
closest approach from the planet: $\dot
x=GM_p/(2Br_p^2\Delta\varphi_s^2)$. That yields:
\begin{equation}
  \tau_{\rm U-turn}=2x_s/\dot x=\Omega_p^2\,h(r_p)^{3/2} q^{-1/2}/(A^2B)\approx
  \frac{4}{\Omega_p}\left[\frac{H(r_p)}{R_H}\right]^{3/2},
\end{equation}
where $R_H=r_p(q/3)^{1/3}$ is the Hill radius of the planet, and where
the last equality holds for a Keplerian disk. When the planet emerges
from the disk ($H(r_p)\sim R_H$), the horseshoe U-turn occurs on the
dynamical timescale. When dealing with an embedded object however
($R_H<H(r_p)$), the horseshoe U-turn time can be substantially longer than
the dynamical time (e.g. $10$ times longer for an Earth mass object
embedded in a disk with $h(r_p)=0.05$).

Using Eq.~(\ref{eq:hslib}), we are led to:
\begin{equation}
\frac{\tau_{\rm lib}}{\tau_{\rm U-turn}}\approx h(r_p)^{-1}.
\end{equation}
There is at least an order of magnitude difference between the
horseshoe U-turn time and the libration time in a thin disk, hence it
should be possible to find a location in the disk where the cooling
time is much longer than the U-turn time and yet shorter than the
libration time, so as to maintain an unsaturated corotation torque.

\subsection{Entropy gradient and baroclinic instability}
\label{sec:baroc}

The effect that we present in this two-dimensional analysis is of 
particular interest when there is a negative entropy gradient at 
corotation, since this may suffice to halt type~I migration. 
It would be of interest to generalize the present analysis to the case 
of a three-dimensional baroclinic disk. We comment also 
that in such systems, a negative 
entropy gradient may render the disk unstable to a baroclinic instability 
\citep{kl04,kb03}. It is certainly important to examine the interplay of the
baroclinic instability and of the corotational effects presented here.
The turbulence generated by the baroclinic instability, in particular,
could provide a mechanism to prevent the saturation of the corotation
torque, much like the turbulence arising from the MRI can prevent
the corotation torque saturation in an isothermal disk.

\section{Conclusions}
\label{sec:conc}

We evaluate the corotation torque between an adiabatic gaseous disk
and a uniformly rotating external potential. In the linear case for an
isolated resonance, we find a singular contribution at corotation
which scales with the entropy gradient, and which arises from the
advection of entropy within the libration region. This effect neither
exists in isothermal or locally isothermal flows, nor does it exist
for barotropic fluids (such as fluids described by a polytropic
equation of state). We provide a torque expression at an isolated
resonance which involves the pressure perturbation at corotation. We
then check the torque expression by two-dimensional adiabatic calculations that
involve an isolated resonance. In particular, we exhibit a case with a
flat vortensity profile, for which the corotation torque does not
cancel out and is in correct agreement with the analytical expression.
We then turn to the case of an embedded planet, for which we find an
excess of corotation torque in the adiabatic case, which scales with the entropy
gradient. For a sufficiently small planet mass, we check that this
excess can be accounted for by a summation over the resonances of the
torque excess that we found in the first part. This confirms that this
effect is essentially a linear effect. We finally discuss in
section~\ref{sec:discu} some open questions linked to the softening
length, to the saturation, to the case of a three-dimensional baroclinic 
disk, and to the interplay with the baroclinic instability, on to which 
theoretical efforts should focus in a nearby future.

\acknowledgements
We wish to thank Sijme-Jan Paardekooper and John C.B. Papaloizou for
interesting discussions on the topics covered in this manuscript. We also 
thank Alessandro Morbidelli for a thorough reading of a first version of this 
manuscript, and an anonymous referee for comments that led to an improvement
of the paper.

\appendix

\section{Additional contribution to the corotation torque }
\label{apA}

\begin{figure}
  \plottwo{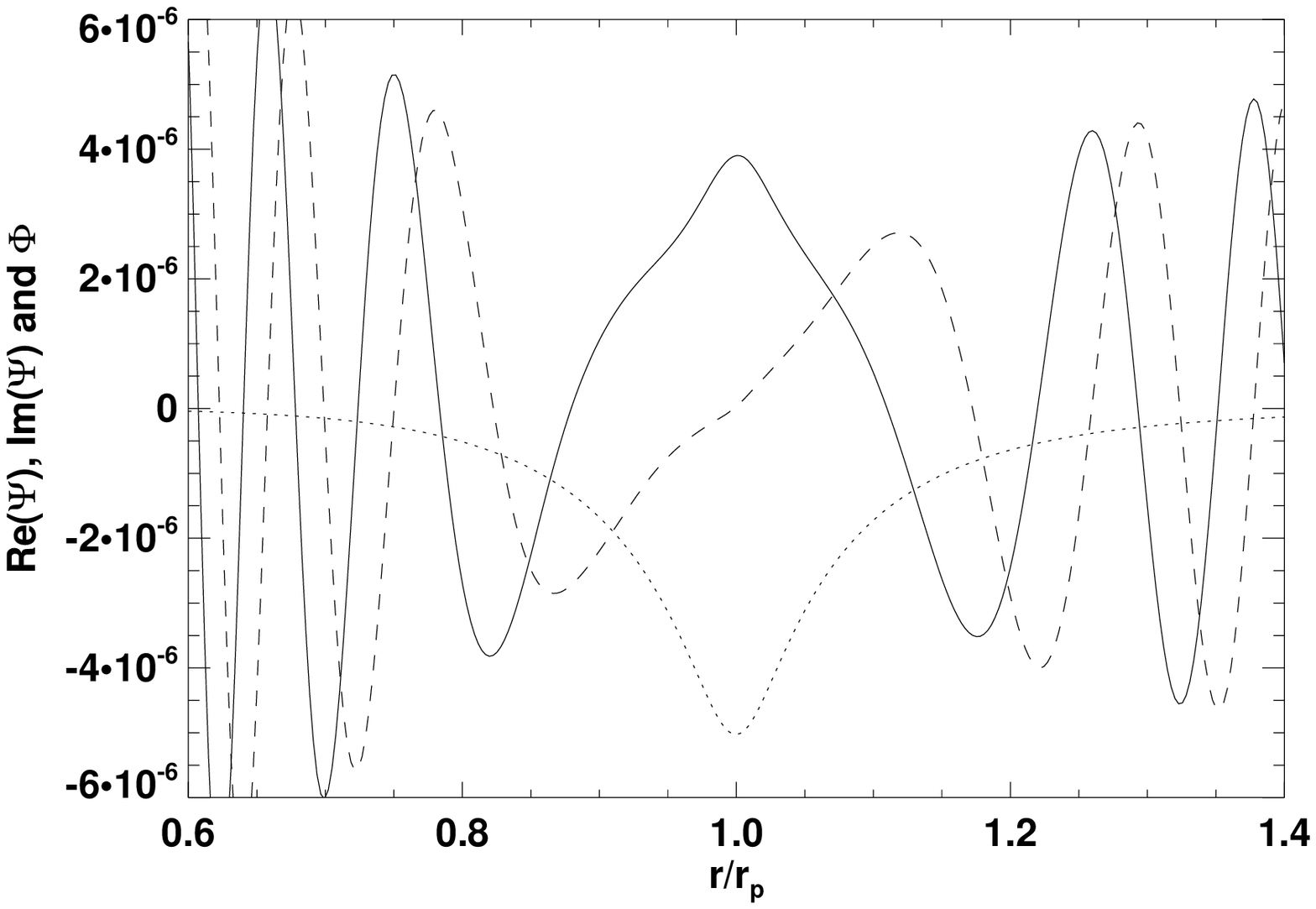}{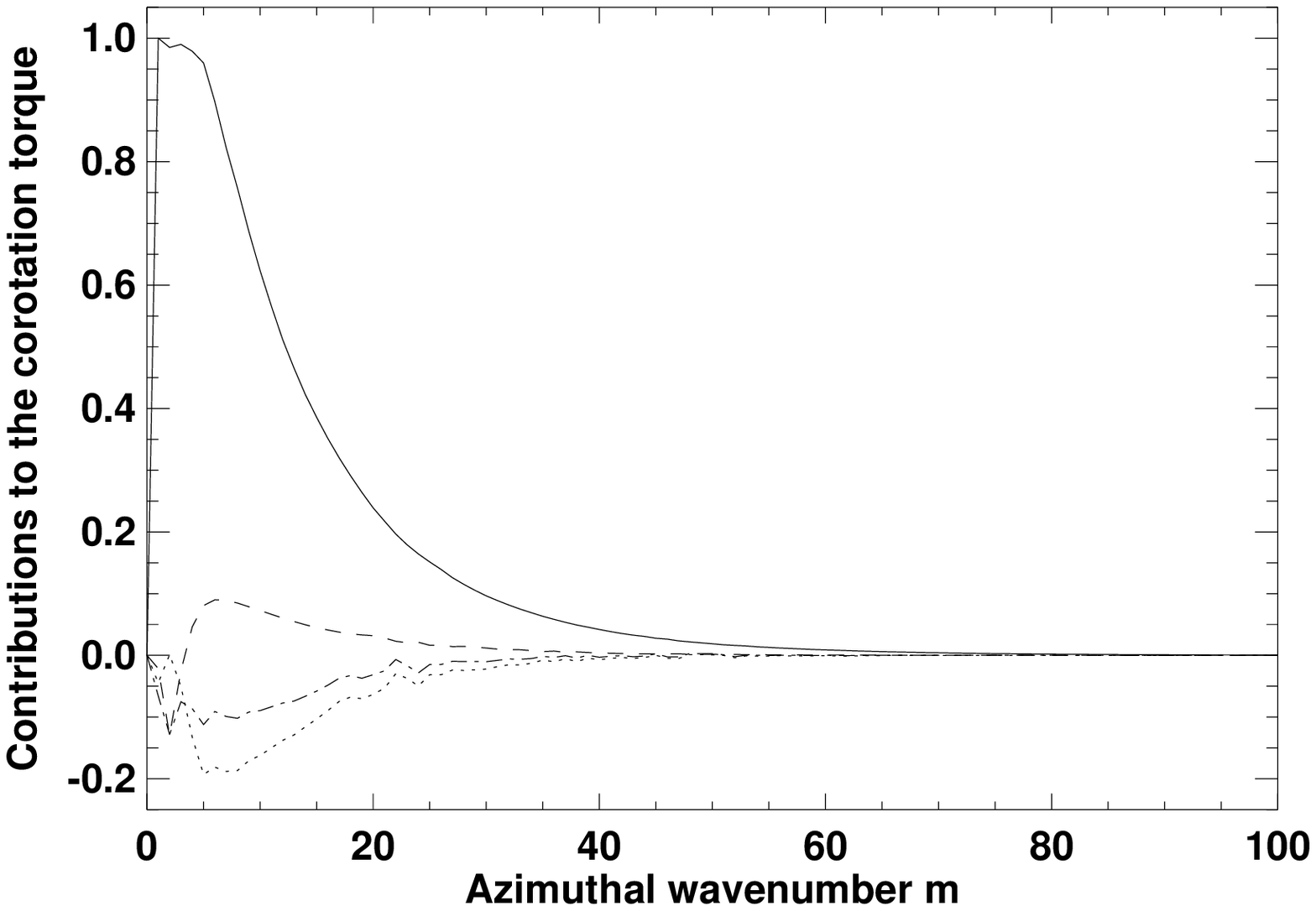} \figcaption{\label{append}Left: $m=8$
    azimuthal component of $\Psi=p/\Sigma_0$ (the real part is shown
    by a solid line, the imaginary part is shown by a dashed
    line). The dotted line shows the $m=8$ component of the potential
    (which is purely real).  Right: singular contribution of $\Psi$ at
    corotation (solid line), contribution of the extra term in
    $\Im[\Psi(x)]/x$ of Eq.~(\ref{ims1approx}) (dotted line),
    contribution of the first term of Eq.~(\ref{s0u1}) (dashed line),
    and total contribution of these extra terms (dash-dotted
    line). Each contribution is normalized to the maximum value
    of the singular contribution of $\Psi$ at corotation.}
\end{figure}

We check hereafter that the contribution of the last term of
Eq.~(\ref{ims1approx}) is negligible for an embedded planet. 
For this purpose, we
compare $G=\int dx\,\Im[\Psi(x)]\Phi(x)/x$ to $-\pi[\Re(\Psi)\Phi]_{r_c}$.
Fig.~\ref{append}a shows the $m=8$ component of $\Psi=p/\Sigma_0$ 
for the calculation presented in 
section~\ref{sec:connect}. We clearly see that the behavior of
$\Im(\Psi)$ in the vicinity of corotation comes from the overlap of
the behavior arising at the inner and outer Lindblad resonances.
Between its outermost inner minimum at $r_-\sim 0.87$ and its
innermost outer maximum at $r_+\sim 1.12$, $\Im(\Psi)$ can be
considered as having a linear dependence in
$x$. Notwithstanding the decrease of $\Phi$ as one recedes from
corotation, the main contribution to $G$ will come from $\Im[\Psi(x)]/x$
between these two radii, as it exhibits a flat behavior over this
range. This yields $G \sim 2\Phi(r_c)|\Im[\Psi(r_\pm)]|$, that it to say a
result comparable in order of magnitude to $-\pi[\Re(\Psi)\Phi]_{r_c}$.
Nevertheless, the final contribution of the extra term is much smaller
than the singular one at corotation for the following reasons:

\begin{enumerate}
\item The potential decreases sharply as one recedes from corotation,
  which provides a cut-off to the extra term, that is not localized at
  corotation.
\item The extra term is partially compensated for by the first term of
  Eq.~(\ref{s0u1}), which we have neglected in writing
  Eq.~(\ref{s1approx}), and which yields another term in
  Eq.~(\ref{ims1approx}) that reads $-({\cal F}{\cal S} /
  r^2\Omega)_{r_c} d\Im[\Psi(x)]/dx$.  Adding this additional term and
  the last term of Eq.~(\ref{ims1approx}), we are left with:
  \begin{eqnarray*}
    \left.
      -\frac{d\Im[\Psi(x)]}{dx}\left[ \frac{{\cal F}{\cal S}}{r^2\Omega}Ê\right]_{r_c}
      -\frac{\Im[\Psi(x)]}{x}\left[\frac{2{\cal FS}}{r^3\Omega'}\right]_{r_c}
    \right.
    &\sim&
    -\frac{\Im[\Psi(x)]}{x} \left\{ \frac{r\Omega^{'}}{2\Omega} + 1 \right\}  
    \left[\frac{2{\cal FS}}{r^3\Omega'}\right]_{r_c}\\
    &\sim& 
    -\frac{\Im[\Psi(x)]}{x}\left[\frac{{\cal FS}}{2r^3\Omega'}\right]_{r_c},
  \end{eqnarray*}
  which shows that, in addition to the potential cutoff, the last term
  of Eq.~(\ref{ims1approx}) should be decreased by a factor of $4$ (we
  neglect, at this level of accuracy, the jump in $\Im(\Psi')$ at
  corotation).
\end{enumerate}
We have checked on the calculation presented in section~\ref{sec:connect} 
that the contribution of these extra terms is indeed small compared to the
singular contribution at corotation at all $m$. This is shown in
Fig.~\ref{append}b, from which we can conclude that the total contribution
of the extra terms is about an order of magnitude smaller than the
singular contribution. The agreement that we found in
section~\ref{sec:connect} between the numerical simulation of an
embedded planet and the torque series, which was of the order of a
percent, might then be fortuitous. Nevertheless, we expect an
agreement of the order of $10$~\%, still very satisfactory. These
findings are also compatible with the fact that we hardly see any
diffuse torque density outside of the horseshoe region in
Fig.~\ref{tqdens}b.

\end{document}